\documentclass[journal = aaembp, manuscript = paper]{achemso}
\setkeys{acs}{maxauthors = 1000}

\usepackage[version=3]{mhchem} 
\usepackage[T1]{fontenc}       
\usepackage{xcolor}
\usepackage{mhchem}
\usepackage[range-units=single]{siunitx}
\usepackage{ulem}
\usepackage{amssymb}
\RequirePackage{epstopdf}
\DeclareGraphicsExtensions{.eps}
\DeclareSIUnit{\wtpercent}{wt.\%}
\DeclareSIUnit{\curr}{\textit{I}}
\DeclareSIUnit{\Temp}{\textit{T}}
\DeclareSIUnit{\Press}{\textit{P}}
\DeclareSIUnit{\volt}{\textit{V}}

\usepackage{hyperref}

\author{Yorick~A.~Birkh\"{o}lzer}
\affiliation[University of Twente]
{MESA+ Institute of Nanotechnology, University of Twente, P.O. Box 217, 7500AE Enschede, The Netherlands}
\alsoaffiliation{Equally contributed}
\email{y.a.birkholzer@utwente.nl}
\author{Kai~Sotthewes}
\affiliation[University of Twente]
{MESA+ Institute of Nanotechnology, University of Twente, P.O. Box 217, 7500AE Enschede, The Netherlands}
\alsoaffiliation{Equally contributed}
\email{k.sotthewes@utwente.nl}
\author{Nicolas~Gauquelin}
\affiliation[University of Antwerp]
{Electron Microscopy for Materials Science (EMAT), University of Antwerp, Groenenborgerlaan 171, 2020 Antwerp, Belgium}
\author{Lars~Riekehr}
\affiliation[University of Antwerp]
{Electron Microscopy for Materials Science (EMAT), University of Antwerp, Groenenborgerlaan 171, 2020 Antwerp, Belgium}
\author{Daen~Jannis}
\affiliation[University of Antwerp]
{Electron Microscopy for Materials Science (EMAT), University of Antwerp, Groenenborgerlaan 171, 2020 Antwerp, Belgium}
\author{Emma~van~der~Minne}
\affiliation[University of Twente]
{MESA+ Institute of Nanotechnology, University of Twente, P.O. Box 217, 7500AE Enschede, The Netherlands}
\author{Yibin~Bu}
\affiliation[University of Twente]
{MESA+ Institute of Nanotechnology, University of Twente, P.O. Box 217, 7500AE Enschede, The Netherlands}
\author{Johan~Verbeeck}
\affiliation[University of Antwerp]
{Electron Microscopy for Materials Science (EMAT), University of Antwerp, Groenenborgerlaan 171, 2020 Antwerp, Belgium}

\author{Harold~J.W.~Zandvliet}
\affiliation[University of Twente]
{MESA+ Institute of Nanotechnology, University of Twente, P.O. Box 217, 7500AE Enschede, The Netherlands}

\author{Gertjan~Koster}
\affiliation[University of Twente]
{MESA+ Institute of Nanotechnology, University of Twente, P.O. Box 217, 7500AE Enschede, The Netherlands}

\author{Guus~Rijnders}
\affiliation{MESA+ Institute of Nanotechnology, University of Twente, P.O. Box 217, 7500AE Enschede, The Netherlands}

\title{High-strain-induced local modification of the electronic properties of \ce{VO2} thin films}


\begin{document}

\begin{tocentry}

\centering
\includegraphics[width=\columnwidth]{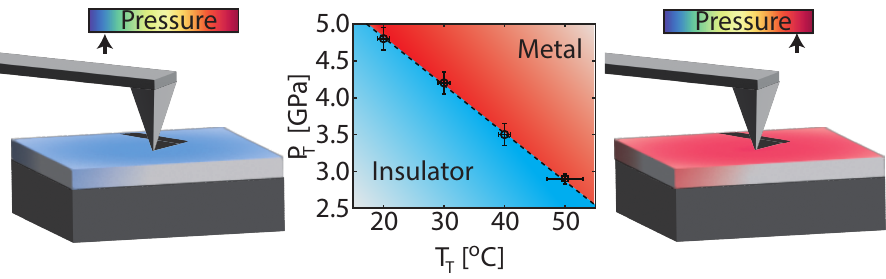}

\end{tocentry}

\doublespacing

\begin{abstract}
Vanadium dioxide (\ce{VO2}) is a popular candidate for electronic and optical switching applications due to its well-known semiconductor-metal transition.
Its study is notoriously challenging due to the interplay of long and short range elastic distortions, as well as the symmetry change, and the electronic structure changes.
The inherent coupling of lattice and electronic degrees of freedom opens the avenue towards mechanical actuation of single domains. 
In this work, we show that we can manipulate and monitor the reversible semiconductor-to-metal transition of \ce{VO2} while applying a controlled amount of mechanical pressure by a nanosized metallic probe using an atomic force microscope.
At a critical pressure, we can reversibly actuate the phase transition with a large modulation of the conductivity.
Direct tunneling through the \ce{VO2}-metal contact is observed as the main charge carrier injection mechanism before and after the phase transition of \ce{VO2}.
The tunneling barrier is formed by a very thin but persistently insulating surface layer of the \ce{VO2}.
The necessary pressure to induce the transition decreases with temperature. In addition, we measured the phase coexistence line in a hitherto unexplored regime. 
Our study provides valuable information on pressure-induced electronic modifications of the \ce{VO2} properties, as well as on nanoscale metal-oxide contacts, which can help in the future design of oxide electronics.
\end{abstract}

\section{Keywords}
C-AFM, VO2, Pressure, Metal-Insulator Transition, Phase Diagram

\section{Introduction}
Electronic phase transitions play a key role in the study of quantum materials and promise versatile applications in the next generation of low-power electronic devices for information processing and storage \cite{delValle2019}, fast optoelectronic switches\cite{Cueff2015}, actuators \cite{Manca2017}, filters \cite{Wan2022}, hydrogen storage \cite{Yoon2016}, and smart windows \cite{Tang2021}. With a phase transition temperature close to room temperature and an electrical resistance change of several orders of magnitude \cite{Morin1959}, vanadium dioxide (\ce{VO2}) is one of the best-studied candidates from the class of transition metal oxides.

The reversible semiconductor-to-metal transition, more often referred to as metal-insulator transition (MIT), can be triggered by various stimuli, such as heat \cite{Le2019}, light \cite{Vidas2020}, mechanical pressure \cite{Bai2015}, electric \cite{delValle2021}, and magnetic fields \cite{Matsuda2020}, which opens the possibility for realizing switching devices between well-defined off (insulating) and on (metallic) states. In order to better understand the microscopic origin of the MIT, \ce{VO2} has been investigated with state-of-the-art (pump-probe) spectroscopy techniques employing radiation all across the electromagnetic spectrum from hard and soft X-rays \cite{Johnson2021}, to XUV \cite{Jager2017}, infrared \cite{Sternbach2021}, \SI{}{\THz} \cite{Otto2019}, microwave \cite{Zinsser2019}, and radio-frequency \cite{Muller2020}; as well as electron microscopy \cite{Sood2021, Fu2020}. The interested reader is referred to the review by Shao \textit{et al.} \cite{Shao2018}. 
Our current understanding is that the electronic phase transition of \ce{VO2} is inherently coupled to an accompanying structural phase transition from a monoclinic crystal structure at low temperature to a tetragonal structure at high temperature. The monoclinic phase is defined by a characteristic dimerization and tilt motif of the vanadium ions, whereas the tetragonal phase is isostructural to rutile \ce{TiO2}. However, despite decades of research, the exact mechanism of the phase transition (Mott \cite{Zylbersztejn1975}, Peierls \cite{Goodenough1971}, order-disorder \cite{Wall2018}) is still under debate.

In this context, the possibility of switching thin film quantum materials by pressure has so far received limited attention. In the case of \ce{VO2}, an externally applied pressure has been used before to induce the MIT \cite{Neuman1964,Berglund1969,Park2013,Chen2017,Baldini2016,Cao2009}. In the work of Park \textit{et al.} \cite{Park2013}, \ce{VO2} single-crystal nanobeams were studied under the influence of uniaxial stress in a scanning electron microscope. From these measurements, a pressure-temperature (\SI{}{\Press}-\SI{}{\Temp}) phase diagram was constructed. To date, the MIT was controlled on a macroscopic scale or by a macroscopic probe, but reliable and reversible nanoscale control of the material and its properties remained elusive. We expect to gain important insights provided that sufficient pressure levels are achieved, and pressure is directly applied in the crystallographic direction where the dimerization takes place. 

There are two main routes to introduce strain or pressure into the system. One is strain engineering such as heteroepitaxy, where a thin film is grown on a crystalline substrate that acts as a template and clamps the in-plane lattice parameter. Epitaxial strain can be anisotropic but is usually biaxial, i.e., applying pressure within the interfacial plane and only indirectly - through the Poisson effect - in the direction of the surface normal. This is in contrast to the second route in which pressure is applied externally by, for instance, a diamond anvil cell (DAC) \cite{Chen2017}. Thanks to their limited contact area in the range of a few hundred \SI{}{\micro\meter} in diameter, DACs can reach uniaxial pressures on the order of hundreds of \SI{}{\GPa}. In a DAC, a pressure medium can be used to convert uniaxial pressure into hydrostatic, isotropic pressure. 

Here, we combine the two routes, (i) the effect of epitaxy, i.e., in-plane clamping, with (ii) the principle of a uniaxial DAC, albeit on an approximately 4 orders of magnitude smaller length scale by using a diamond AFM tip. As we will show, this allows manipulation and symmetry breaking of the material under study. 
Atomic force microscopy (AFM) is a very suitable method for applying large pressures (up to tens of \SI{}{\GPa}) on various types of materials. To date, nanoscale manipulation with an AFM has been performed on \ce{NiO} \cite{Kim2013}, \ce{Sr2IrO4} \cite{Domingo2015}, \ce{SrTiO3} \cite{Das2019}, \ce{V2O3} \cite{Alyabyeva2018}, \ce{VO2} \cite{Schrecongost2019}, and two-dimensional ice \cite{Sotthewes2017}. In this way, the material can locally be manipulated while, for instance, the transport mechanism can be determined simultaneously from nanoscale current-voltage (\SI{}{\curr}(\SI{}{\volt})) spectroscopy. 
For more background on nanoelectrical AFM measurements, the interested reader is referred to the books by \citet{Celano2019} and \citet{Lanza2017}.
Schrecongost \textit{et al.} \cite{Schrecongost2019} were able to demonstrate (irreversible) local manipulations of \ce{VO2} under very large electric fields. By applying probe-induced strain or by a highly biased AFM probe, a persistent metallic phase or an insulating phase could be created through plastic deformation and electrochemical modification. 

In the present study, we apply an external, local, and predominantly uniaxial stress on an epitaxially grown and biaxially pre-strained \ce{VO2} thin film using a conductive AFM tip in order to probe the reversible metal-insulator transition (MIT) in a fully elastic regime without changing the stoichiometry. Furthermore, we elucidate the electrical contact mechanism and expand the \SI{}{\Press}-\SI{}{\Temp} phase diagram by an order of magnitude with respect to the seminal results of Park \textit{et al.} \cite{Park2013}.

\section{Results and discussion}
\begin{figure}[t!]
	\centering
	\includegraphics[width=\linewidth]{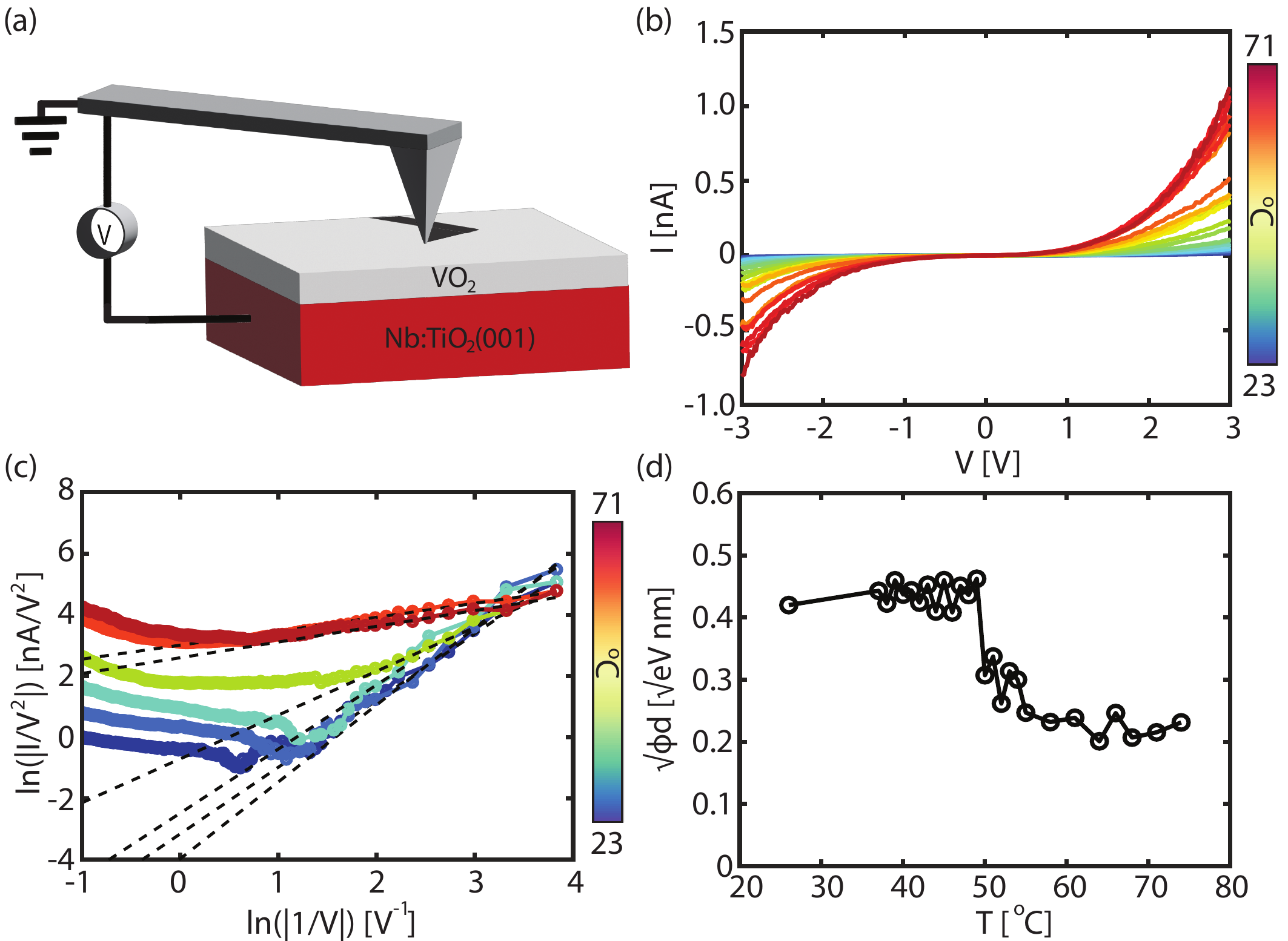}
	\caption{(a) Schematic representation of the experimental setup. 
	The voltage is applied between the tip and the \ce{Nb}:\ce{TiO2} substrate. 
	(b) The median \SI{}{\curr}(\SI{}{\volt}) curves recorded at different temperatures. 
	(c) The same data as in (b) plotted on a $\ln(I/V^2)$ versus $\ln(|1/V|)$ scale (only the negative polarity is shown), which is related to the direct tunneling (DT) model. 
	The black dashed lines are fits using equation \ref{DT-model}. 
	(d) The obtained barrier height parameters ($\sqrt{\phi_\textrm{B}}\,d$) from panel (c) versus the temperature (\SI{}{\Temp}). 
	Around \SI{50}{\degreeCelsius}, a clear transition is observed, which is related to the metal-insulator transition in \ce{VO2}.}
\label{Fig_Temp_dep_mes}
\end{figure}

\SIrange[]{9}{15}{\nm} thick \ce{VO2} thin films were grown on \SI{0.5}{\wtpercent} \ce{Nb}-doped \ce{TiO2} (001) substrates (\ce{Nb}:\ce{TiO2}) by pulsed laser deposition (see experimental section and the Supp. Mat. for sample preparation and characterization details). This orientation is chosen such that the crystallographic axis of \ce{VO2} along which the dimerization takes place in the low-temperature phase (c-axis in the rutile high-temperature phase) points in the out-of-plane direction.
The electronic properties are determined by placing a highly boron-doped, single crystalline diamond (BDD) AFM tip in direct contact with the \ce{VO2} film. 
The amount of load on the sample is controlled by the force feedback, while the current flows between the tip and the \ce{Nb}:\ce{TiO2} bottom electrode as a function of the applied voltage (\SI{}{\volt}) (see schematic in \autoref{Fig_Temp_dep_mes}\,a). 
The tip acts as the top electrode as well as a nanoscale mechanical indenter. 
Although the applied pressures are going as high as \SI{8}{\GPa}, the mechanical load is kept below the plastic deformation threshold of the material to avoid permanent deformation or damage to the \ce{VO2} film (see Supp. Mat.). 

First, the contact properties of the BDD/\ce{VO2}/\ce{Nb}:\ce{TiO2} junction are measured as a function of temperature (\SI{}{\Temp}). 
The applied pressure during the experiment is kept constant at approx. \SI{2.7}{\GPa}. 
Local transport measurements reveal a gradual change of the obtained current (\SI{}{\curr}) versus voltage (\SI{}{\volt}) curves with increasing \SI{}{\Temp} (see \autoref{Fig_Temp_dep_mes}\,b). 
Because of the semiconducting nature of \ce{VO2} at room temperature, a Schottky barrier ($\phi_\textrm{B}$) is formed between the conducting tip and the \ce{VO2} film. 
In general, the Schottky barrier is dependent on the metal work function ($\phi_\textrm{M}$) and the electron affinity of the semiconductor ($\chi$) \cite{Tung2014,Bampoulis2017}. 
However, phenomena like Fermi level pinning can cause a deviation from the standard Schottky-Mott rule \cite{Tung2014,Sotthewes2019}. 
In our experimental structure, a nanoscale (top) electrode is used (AFM tip), and a second macroscopic (bottom) electrode is required (\ce{VO2}/\ce{Nb}:\ce{TiO2} interface) to close the electrical circuit. 
Both electrodes are typically described as two Schottky diodes reversibly connected in series \cite{Nouchi2014}. 
Because the tip/\ce{VO2} contact is much smaller in size compared to the macroscopic contact (\num[]{9} orders of magnitude), the current blockage by the macroscopic contact is negligible and the nanoscopic contact dominates the current flow. 
Therefore, our circuit can effectively be described as a single metal-semiconductor junction \cite{Bampoulis2017}.

To explain the evolution of the \SI{}{\curr}(\SI{}{\volt}) curves as a function of temperature, the Schottky barrier height is extracted. 
There are three carrier injection mechanisms that are dominant in the current flow across a Schottky barrier \cite{DuranRetamal2018,Bremen2019}.
Firstly, there is thermionic emission; secondly, there is direct tunneling (DT), and thirdly, there is Fowler-Nordheim (F-N) tunneling. These mechanisms are influenced by the applied bias configuration, doping level, temperature, and the exact interface composition. 
In most cases, the carrier injection mechanism across a metal-semiconductor interface is thermionic emission. 
From the thermionic emission model, both the Schottky barrier height $\phi_\textrm{B}$ and the ideality factor $\eta$ can be extracted (see Supp. Mat. for an elaborate explanation and analysis). 
Both quantities  are dependent on the temperature (as shown in Figure S10 in Supp. Mat.). For instance, $\phi_\textrm{B}$ increases with increasing temperature while $\eta$ is decreases. 
The ideality factor is typically used to assess the deviation of the current transport from ideal thermionic emission, where $\eta < 2$ implies pure thermionic emission. 
These trends have  also been observed for other material systems, such as for metal electrodes on silicon \cite{Werner1993}. 
At lower temperatures, the charge carriers have insufficient energy to surmount the Schottky barrier. 
Therefore, the current transport is dominated by other carrier injection mechanisms, resulting in higher ideality factors. 
With increasing temperature, more charge carriers gain sufficient energy to overcome the Schottky barrier, resulting in a lowering of the ideality factor. 
Around \SI{50}{\degreeCelsius}, a clear change in the slope of the ideality factor is observed. 
This point coincides with the metal-insulator transition temperature ($T_\textrm{MIT}$) obtained from the macroscopic transport measurement (see Figure S15 in Supp. Mat.).

The large ideality factors obtained from \autoref{Fig_Temp_dep_mes}\,b ($\eta >>$ 2) (details are shown in Figure S10-S12 in the Supp. Mater.) imply that other charge injection mechanisms beyond thermionic emission play an important role. 

The DT current depends linearly on the bias according to \cite{Beebe2006,Ikuno2011},

\begin{equation}
I = \frac{A_{\textrm{eff}}\,q^2\,V\sqrt{m_0\,\phi_\textrm{B}}}{h^2\,d} \exp\Big[\frac{-4 \pi d \sqrt{m_0\phi}}{h}\Big]
\label{DT-model}
\end{equation}
\\ where $A_\textrm{eff}$ is the effective contact area of the tip/\ce{VO2} junction, $m_0$ the free electron mass, $q$ the electronic charge, $h$ the Planck constant,$d$ the barrier width and $\phi$ the height of the tunnel barrier.

F-N tunneling is described by \cite{Beebe2006,Ikuno2011}

\begin{equation}
I = \frac{A_{\textrm{eff}}\,q^3\,V^2\,m_0}{8\,\pi\,h\,\phi\,d^2 \,m^*} \exp\Big[\frac{-8\,\pi\,d\,\phi^{\frac{3}{2}} \sqrt{2m^*}}{3\,h\,q\,V}\Big]
\label{FN-model}
\end{equation}
\\ where $m^*$ is the effective mass of an electron ($\approx$ 3.4 $m_0$\cite{Muraoka2018}).  

Equations \ref{DT-model} and \ref{FN-model} imply that DT and F-N differ in terms of \SI{}{\curr}(\SI{}{\volt}) dependency. 
Both mechanisms can easily be distinguished when plotting $\ln(I/V^2$) as a function of $1/V$, a so-called F-N plot. 
For F-N, the curve should decrease linearly, whereas for DT, it should increase logarithmically \cite{Beebe2006,Ikuno2011,Bremen2019}. 
DT shows a linear behavior when plotted on a $\ln(I/V^2)$ versus $\ln(|1/V|)$ scale. 
For DT and F-N, it is possible to determine the barrier parameters $\phi^{3/2}\,d$ and $\sqrt{\phi}\,d$, respectively. 
Both parameters are expressed in terms of the barrier width $d$ because the exact width of the barrier is unknown. 
From the F-N plot (see Figure S10 in Supp. Mat.) it is clear that DT is the dominant charge injection mechanism and that F-N plays a minor role in the charge transport here. 
The inflection point \cite{Beebe2006,Ikuno2011,Bremen2019}, which is defined as the point where the two regimes are separated, is located at relatively high voltages ($V\,>\,\SI{2.5}{\volt}$). 
Within this study, the F-N regime cannot be reached as voltages larger than \SI{3}{\volt} permanently degrade the material \cite{Schrecongost2019}.

In \autoref{Fig_Temp_dep_mes}\,c, the data from \autoref{Fig_Temp_dep_mes}\,b is plotted on a $\ln(I/V^2)$ versus $\ln(|1/V|)$ scale, also known as DT plot. 
A clear linear regime is observed, which can be fitted with equation \ref{DT-model} to extract the DT barrier parameters. 
The extracted barrier parameter $\sqrt{\phi}\,d$ is plotted as a function of temperature in \autoref{Fig_Temp_dep_mes}\,d. 
A clear transition is observed around \SI{50}{\degreeCelsius}. 
Before the transition, $\sqrt{\phi}\,d$ is approximately constant, which is expected as tunneling is a temperature-independent process \cite{Karthaeuser2011,Sotthewes2014}. 
The average value \SI{0.40}{\electronvolt^{1/2}.\nm} agrees well with other measurements performed on semiconductors \cite{Bremen2019}. 
Note here, that the total current through the junction is still temperature-dependent, as the contribution from thermionic emission is increasing with temperature (decreasing $\eta$, see Figure S10 in Supp. Mat.).

Surprisingly, after the metal-insulator transition, the barrier height parameter ($\sqrt{\phi}\,d$) still has a finite value ($\sqrt{\phi}\,d\,\approx\, \SI{0.2}{\electronvolt^{1/2}}$). 
For a metal-metal contact, no barrier should be present, and the corresponding \SI{}{\curr}(\SI{}{\volt}) curve should be metallic, i.e., linear. 
The red curves in \autoref{Fig_Temp_dep_mes}\,b clearly deviate from linear behavior. 
As mentioned in the beginning, for most thin films and bulk surfaces, the dominant current injection mechanism is thermionic emission \cite{Nienhaus2007,Hussin2013,Bampoulis2017,Sotthewes2019}, which is contrary to DT being the dominant injection mechanism in our \ce{VO2} thin films. 
In general, DT and F-N are the dominant current injection mechanisms in nanosheets or two-dimensional materials \cite{Ikuno2011,Ahmed2015,Li2015,DuranRetamal2018,Bremen2019}.

Recent X-ray photoemission (XPS) studies revealed that the surface layer of \ce{VO2} contains V ions in a 5+ oxidation state, which is larger than the 4+ oxidation state that is expected for the bulk of stoichiometric \ce{VO2} \cite{Quackenbush2015}. 
Wahila \textit{et al.} furthermore report the absence of a structural phase transition at the surface of \ce{VO2} (001) based on low energy electron diffraction measurements \cite{Wahila2020}. 
For simplicity, we use the common rutile crystallographic basis for both the substrate and the thin film as it applies to the latter above the phase transition temperature. 
Wagner \textit{et al.} \cite{Wagner2021} have studied the surface of \ce{VO2} (110) in great detail and report an oxygen-rich reconstruction of the \ce{VO2} surface reminiscent of \ce{V2O5}, a band insulator that does not possess a metal-insulator transition. As the rutile (110) surface is actually the lowest energy surface of \ce{VO2}, we suspect that similar oxidation also happens at the other higher energy, and hence less-stable, surfaces.

To test this hypothesis, we performed XPS and laboratory-source hard X-ray photoemission spectroscopy (HAXPES) measurements with different takeoff angles on our \ce{VO2} sample.
In phase pure \ce{VO2}, we expect an oxidation state of 4+ for all vanadium ions. 
However, previous reports have shown that the surface of \ce{VO2} is more accurately described by an oxidation state of 5+ due to spontaneously-formed, oxygen-rich surface reconstructions. \cite{Quackenbush2015, Lu2018, Paez2020, Wagner2021}.

\begin{figure}[t!]
	\centering
	\includegraphics[width=\columnwidth]{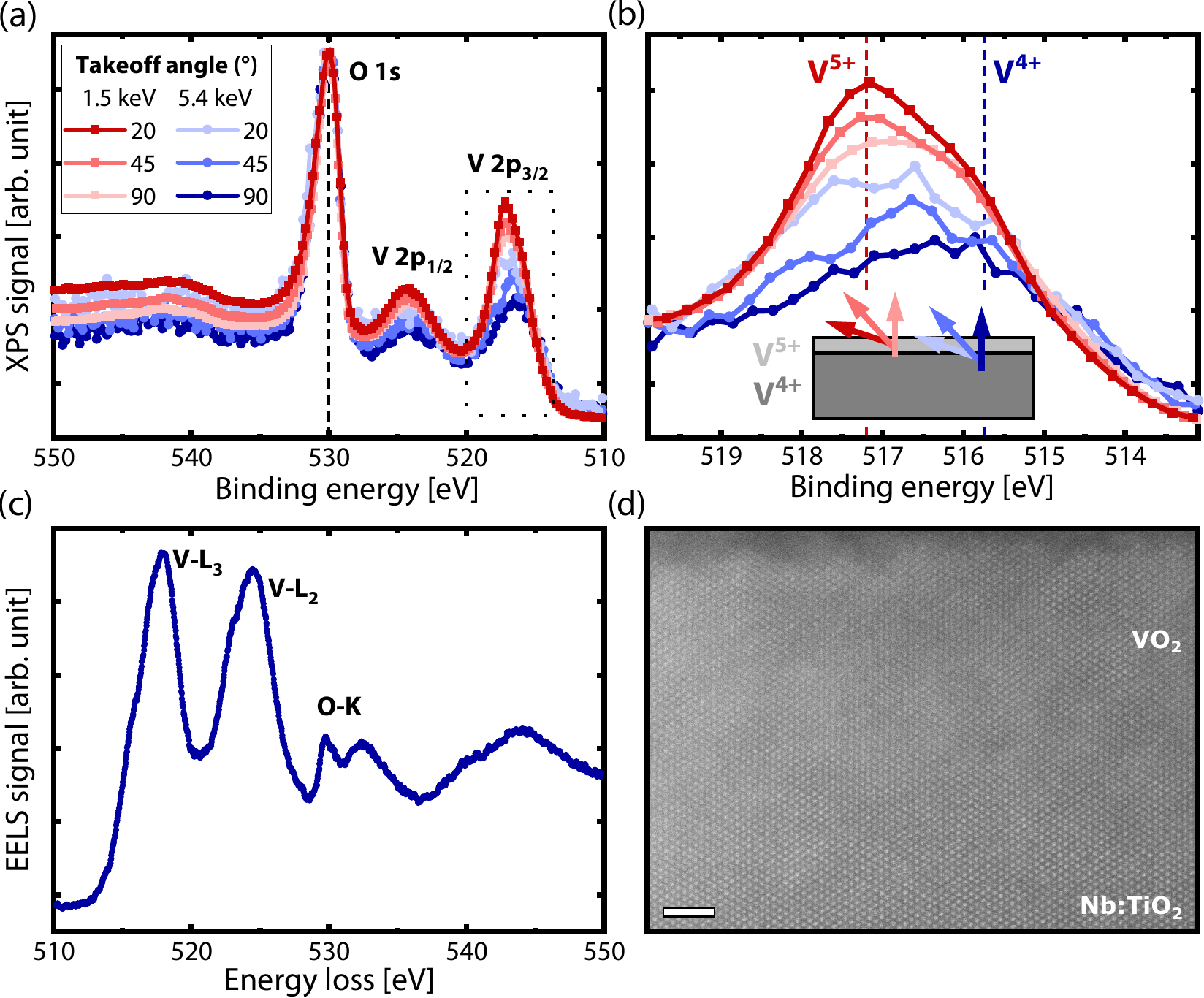}
	\caption{X-ray photoemission measurements using two different photon energies and three different takeoff angles. The inset in panel (b), which is a zoom of the dotted area in panel (a), shows how the takeoff angle is defined with respect to the sample surface. Indicated with red and blue dashed lines are the expected peak positions for \ce{V^{5+}} (\SI{517.2}{\electronvolt}) and \ce{V^4+} (\SI{515.74}{\electronvolt}) according to Silversmit \textit{et al.} \cite{Silversmit2004}. At the surface, the \ce{V} valence is higher (5+) than in the bulk (4+). (c) Electron energy loss spectroscopy (EELS) signal of the interior of the \ce{VO2} layer. (d) Atomic resolution scanning transmission electron microscopy image of the sample cross-section in the [100] zone. The length of the scale bar is \SI{2}{\nm}.}
	\label{Fig_HAXPES}
\end{figure}

We summarize our XPS/HAXPES results in \autoref{Fig_HAXPES}.
At the surface, we find a 5+ oxidation state for vanadium indeed, fully in agreement with the earlier reports mentioned.
By increasing the probing depth of our experiment, which we achieve by increasing the incident photon energy (switching from an \ce{Al} anode to \ce{Cr}) and increasing the photoelectron takeoff angle, the \ce{V} $2p$ peaks shift towards lower binding energies (see \autoref{Fig_HAXPES}\,a). 
Such a shift is characteristic of a reduced oxidation state.
The expected location of the \ce{V} $2p_{3/2}$ peak for \ce{V^5+} is at a binding energy of \SI{517.2}{\electronvolt}, whereas for \ce{V^4+}, it is reduced by \SI{1.46}{\electronvolt} to \SI{515.74}{\electronvolt} \cite{Silversmit2004}.
As we indicated by the dashed reference lines of \autoref{Fig_HAXPES}\,a and b, the magnitude and the direction of the peak shift are consistent with the proposed model.
In the model, it is assumed that the interior of the \ce{VO2} film has the expected vanadium oxidation state of 4+, and the surface has a higher 5+ oxidation state.
We pictorially summarize this observation in the cartoon inset in the bottom panel of \autoref{Fig_HAXPES}\,b.

Unlike tetravalent vanadium oxide (\ce{VO2}), pentavalent vanadium oxide (\ce{V2O5}) is a band insulator that does not possess a metal-insulator transition. \ce{V2O5} has a band gap of \SIrange[]{2.2}{2.4}{\electronvolt} due to the empty $3d$ orbital \cite{Kenny1966}. 
We conclude that it is this pentavalent, insulating surface layer that leads to the tunneling-dominated \SI{}{\curr}(\SI{}{\volt}) characteristics observed by C-AFM. Using this information as input, we can estimate the magnitude of the DT barrier parameter. 
The value for the energy barrier $\phi$ is typically assumed to be half of the band gap of the insulating layer, here $\sim$ \SI{1.2}{\electronvolt}, and we can reasonably assume the thickness of the insulating surface layer to be on the order of one unit cell of \ce{VO2} (001), i.e., $\sim$ \SI{0.28}{\nm}.
The product of the square root of the energy barrier multiplied by the tunneling distance is then $\sim$ \SI{0.3}{\electronvolt^{1/2}.\nm}. This is well in accordance with the experimentally obtained DT barrier parameters shown in \autoref{Fig_Temp_dep_mes} (d).

The insulating \ce{V^5+} surface layer remains present after the semiconductor-to-metal transition of the interior of the \ce{VO2} film and dominates the charge injection mechanism before and after the semiconductor-to-metal transition. However, the structure and electronic properties of the single unit cell layer heavily depend on the structure of the \ce{VO2} thin film. First, in the semiconducting (monoclinic) phase, a tunnel barrier is present in series with a Schottky barrier. For the metallic (rutile) phase, the Schottky barrier vanishes due to the high free carrier density, and only the tunnel barrier is present, significantly lowering the barrier width $d$. Second, most likely, the small shifts of the atomic positions in the \ce{VO2} lattice also lead to small changes in the structure and electronic properties of the pentavalent vanadium oxide surface layer. Analogous effects are also observed for other thin layers \cite{Zheng2010}. These two effects both influence the barrier height parameter of the tunneling process, and therefore a transition is observed in the current-voltage spectroscopy.

Our photoemission study was performed on samples that were exposed to the ambient air just like the samples studied by C-AFM, i.e., \textit{ex situ}. 
We would like to note that even in samples that are transferred in ultrahigh vacuum from the PLD growth to the XPS analysis chamber without exposure to ambient air (\textit{in situ}), we still observe a significant \ce{V^5+} contribution to the expected \ce{V^4+} signal. 
Details on the XPS/HAXPES experiments are provided in the Supp. Mater.

To further investigate the interface between the substrate and the thin film and to confirm the oxidation state of vanadium, we performed scanning transmission electron microscopy (STEM) and electron energy loss spectroscopy (EELS). 
A cross-sectional lamella was prepared by focused ion beam (FIB) milling.
In addition to the position of the \ce{V} $L_{2,3}$ edge, another indicator for the oxidation state and the monoclinic phase of \ce{VO2} at room temperature, i.e., below the phase transition, is the fine structure of the \ce{O} $K$-edge \cite{Abbate1991}.
In \ce{VO2}, the latter is composed of three features, namely the overlapping $\pi^*$ and $d_\parallel^*$ states that form the asymmetric peak around \SI{530}{\electronvolt} and the $\sigma^*$ states that follow around \SI{532.5}{\electronvolt} energy loss.
Away from the top surface, which may have been damaged by the FIB processing, we find a clear \ce{V^4+} character for the interior of the \ce{VO2} thin film, see \autoref{Fig_HAXPES}\,c. This is in full agreement with earlier reports by Tashman \textit{et al.} \cite{Tashman2014} and Lu \textit{et al.} \cite{Lu2020}.
In accordance with the XRD analysis, no crystalline phases other than the expected, commensurately strained \ce{VO2} are observed. This is illustrated by the atomic resolution high-angle annular dark-field (HAADF) image displayed in \autoref{Fig_HAXPES}\,d, which evidences the epitaxial registry of the thin film and the substrate.
Further details on the STEM-EELS analysis and an EELS line profile are provided in the Supp. Mater.

A brief recap, the metal-insulator transition is observed in \ce{VO2} films using a conductive AFM tip while varying the temperature. 
Even when the interior of the film becomes metallic, the interface layer acts as a small barrier between the conductive tip and the \ce{VO2} thin film, resulting in direct tunneling as the dominant transport mechanism.

Besides monitoring the change of the contact mechanism with a varying temperature at a constant pressure, a similar experiment can be conducted by varying the applied pressure at a constant temperature. 
In this case, the load applied by the AFM tip on the \ce{VO2} surface is varied, and simultaneously the \SI{}{\curr}(\SI{}{\volt}) characteristics are measured. 
This enables the opportunity to measure the current injection mechanism under different applied pressures, including the metal-insulator transition. 
Furthermore, rastering allows for recording spatial maps and subsequent averaging over local differences in topography.

\begin{figure}[tb]
	\centering
	\includegraphics[width=\columnwidth]{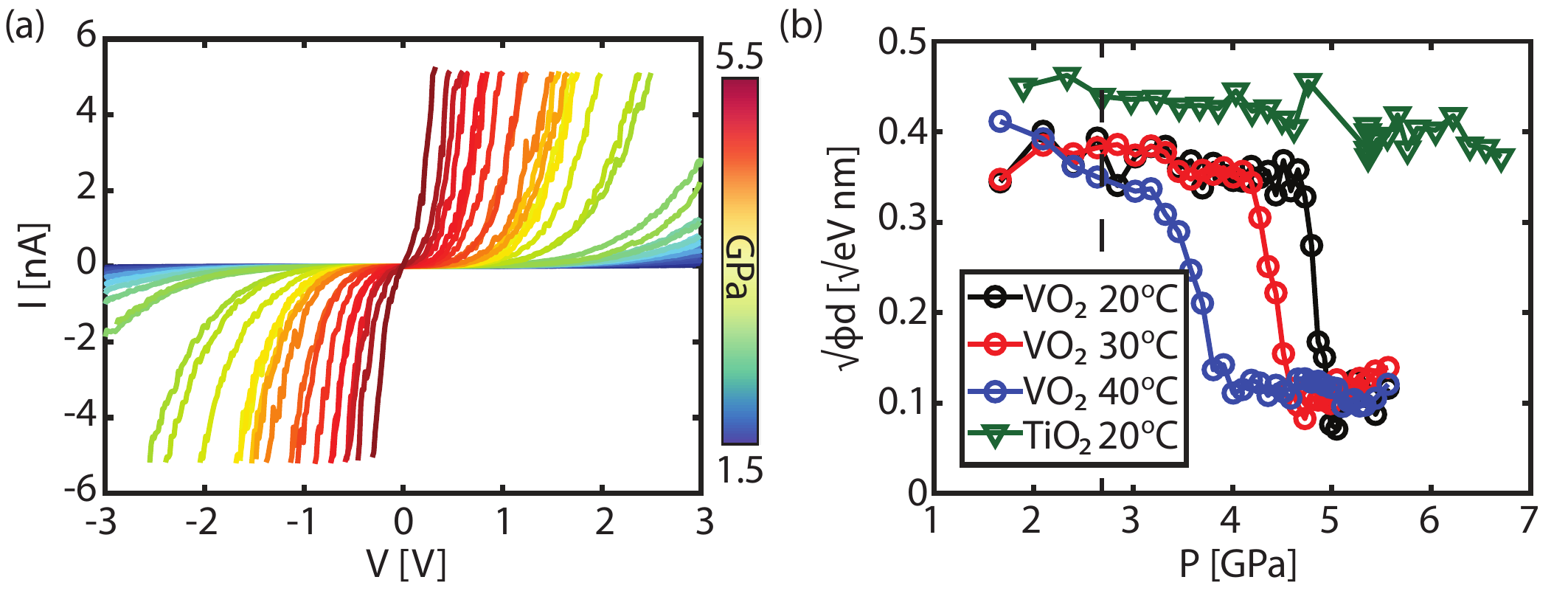}
	\caption{(a) \SI{}{\curr}(\SI{}{\volt}) curves recorded with a doped diamond tip on \ce{VO2} under different applied pressures at \SI{50}{\degreeCelsius}. (b) The obtained barrier height parameter ($\sqrt{\phi_\textrm{B}}\,d$) using equation \ref{DT-model} versus the applied pressure (\SI{}{\Press}) measured at different temperatures (black = \SI{20}{\degreeCelsius}, red = \SI{30}{\degreeCelsius}, blue = \SI{40}{\degreeCelsius}). A clear transition is observed for all three temperatures. The green line is a reference measurement of \ce{TiO2}, which is a material with no metal-insulator transition. The dashed black line is the applied pressure for the varying temperature measurement in \autoref{Fig_Temp_dep_mes}\,d.}
\label{Fig_Pres_var}
\end{figure}

In \autoref{Fig_Pres_var}\,a, the evolution of the \SI{}{\curr}(\SI{}{\volt}) characteristics for a BDD/\ce{VO2}/\ce{Nb}:\ce{TiO2} junction at \SI{50}{\degreeCelsius} is shown for pressures varying between \SI{1.5}{\GPa} and \SI{5.5}{\GPa}. 
Similar to the varying temperature experiment, an increase in current is observed for increasing pressures. 
As a reference measurement, the experiment is repeated on a conventional semiconductor (\ce{TiO2}, see Supp. Mat. for sample preparation and characterization details). 
Similar to \ce{VO2}, the obtained ideality factor is large ($\eta$ > 4), indicating that other charge injection mechanisms are more dominant compared to the thermionic emission model. 
Using the F-N plot, we find that the inflection point is absent, indicating that direct tunneling is again the dominant charge injection mechanism at the used bias voltages. 
The extracted barrier height parameter is plotted as a function of pressure for \ce{TiO2} (Green triangles in \autoref{Fig_Pres_var}\,b). 
A monotonic decrease of the barrier height parameter is observed. 
As the barrier height parameter is a function of the product of the barrier height ($\phi$) and the barrier thickness ($d$), and both parameters are affected by the applied pressure, the exact influence of either parameter cannot be individually analyzed. 
For instance, the band gap of the material may change with applied pressure \cite{Xiao2017,Zhang2018}, affecting $\phi$, while $d$ is also reduced with increasing pressure. 
However, most likely, the reduction of $d$ is the main mechanism behind the reduction of $\sqrt{\phi_\textrm{B}}\,d$ in \ce{TiO2}, as the tunneling current is exponentially dependent on $d$. 
In addition, no abrupt change in the barrier height parameter is observed as a function of pressure in the case of \ce{TiO2}.

Besides \ce{TiO2}, the barrier height parameter ($\sqrt{\phi}\,d$) is also obtained for \ce{VO2} as a function of pressure for different temperatures, see \autoref{Fig_Pres_var}\,b. 
In contrast to \ce{TiO2}, a clear transition is observed in $\sqrt{\phi}\,d$, which is dependent on the temperature. 
At low applied pressures, $\sqrt{\phi}\,d$ is decreasing slightly with pressure. 
Most likely, this reduction is caused by the small decrease in the barrier width $d$, similar to \ce{TiO2}. 
At higher pressures ($\SI{}{\Press} > \SI{3}{\GPa}$), however, an abrupt transition is observed.  
The observed transition as a function of pressure is very similar to the transition observed within the varying temperature experiment (see \autoref{Fig_Temp_dep_mes}), except for the absolute $\sqrt{\phi}\,d$ values of the barrier parameter after the MIT. 
The barrier parameters after the MIT are \SI{0.2}{\electronvolt^{1/2}.\nm} and \SI{0.1}{\electronvolt^{1/2}.\nm} for a varying temperature and pressure, respectively. 
The discrepancy in the barrier parameter reveals that both temperature and pressure have a different influence on the contact mechanism. 
As the temperature is increased, the structure and thereby also the electronic properties of the surface layer change. 
Under applied pressure, however, additional compression is introduced into the interfacial layer, reducing the barrier width even more. 
In other words, an increase in the pressure not only induces the MIT, but also reduces the effective barrier width. 

\begin{figure}[tb]
	\centering
	\includegraphics[width=0.6\columnwidth]{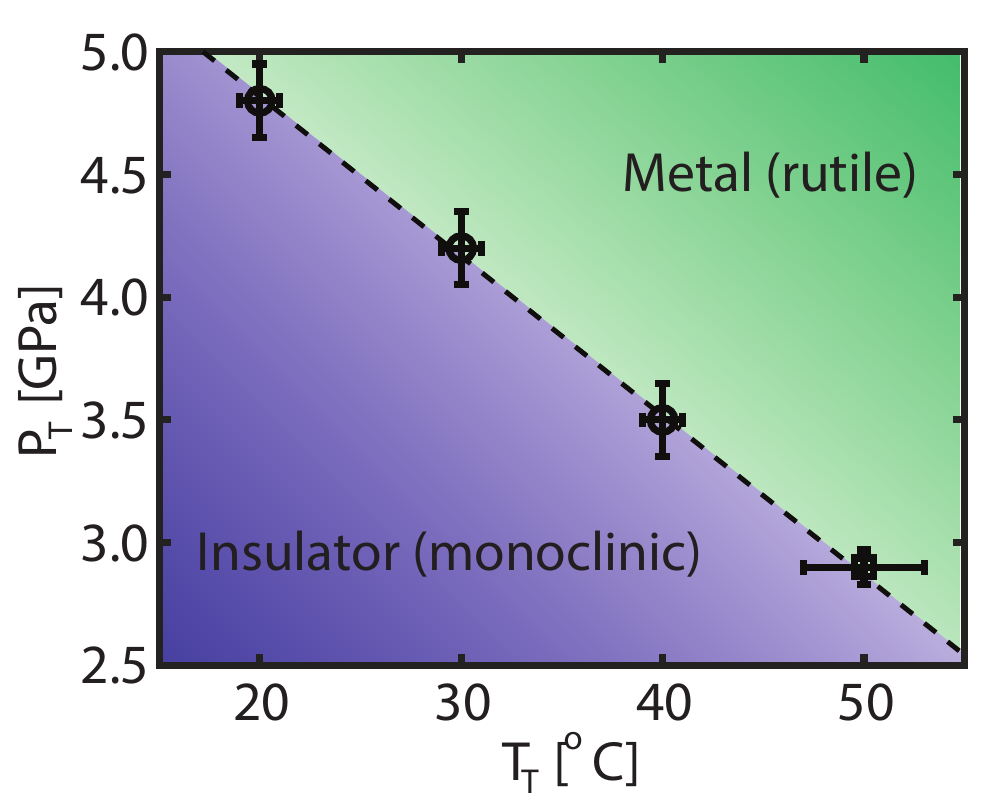}
	\caption{The transition pressure ($P_\textrm{T}$) versus the transition temperature ($T_\textrm{T}$). The measurement points are extracted from the varying temperature (\autoref{Fig_Temp_dep_mes}, $\bigcirc$) and varying pressure (\autoref{Fig_Pres_var}, $\square$) experiment.}
\label{Fig_Phase_diagram}
\end{figure}

When the temperature of the \ce{VO2} is increased, the applied pressure required to induce the metal-insulator transition decreases (see \autoref{Fig_Pres_var}\,b). 
From the varying temperature (\autoref{Fig_Temp_dep_mes}) and pressure (\autoref{Fig_Pres_var}) experiment, the transition pressure ($P_\textrm{T}$) and the transition temperature ($T_\textrm{T}$) are extracted. 
In \autoref{Fig_Phase_diagram} the transition pressure ($P_\textrm{T}$) as a function of the transition temperature ($T_\textrm{T}$) is shown.

Along the coexistence line (dashed line in \autoref{Fig_Phase_diagram}), both phases exist at a constant pressure and temperature. 
Using the Gibbs-Duhem equation, one can derive the Clapeyron relation

\begin{equation}
\frac{\text{d}P}{\text{d}T} = \frac{\Delta S}{\Delta V}
\label{Gibbs-Duhem}
\end{equation}
\\ with $\Delta S$ the entropy and $\Delta V$ the volume changes across the solid-solid phase transition. The entropy per volume ratio can be determined from the slope of the coexistence line. From the slope (linear regression), a value of \SI{60}{\MPa.\degreeCelsius^{-1}}  is found, which is in excellent agreement with previously reported values for macroscopic measurements on the phase transition of \ce{VO2}. \cite{Berglund1969a,Park2013}

What makes pressure-controlled C-AFM exciting for the study of quantum materials is the enormous range of pressures that can conveniently and routinely be accessed under the sharp probe apex and the resulting giant changes in the local structural and electronic properties of the material this can unleash. 
Assuming a Young's modulus of \SI{128}{\giga\pascal} \cite{Guo2011}, we linearly convert stress into strain using Hooke's law in its one-dimensional scalar form. The transition pressures of \SIrange[]{3}{5}{\giga\pascal} displayed in \autoref{Fig_Phase_diagram}, which are needed to drive the MIT of \ce{VO2} by nano-indentation, then correspond to vast uniaxial compressive strains along the rutile c-axis of approx. \SIrange[]{2}{4}{\percent}. 
Although the exact pressure distribution in AFM-based nano-indentation is more complicated than one-dimensional compression, we confirmed with finite element modeling in the Supp. Mater. that the uniaxial out-of-plane compression term dominates over lateral effects and indeed extends throughout the thickness of our films.

Strains of several percent are roughly one order of magnitude more than what bulk oxide specimens can usually withstand before cracking and thus greatly expand the iconic phase diagram of Park \textit{et al.}\cite{Park2013}. 
In addition, when down-scaling electronics towards the nanoscale, the contacts become increasingly more important. 
Using an AFM tip as an electrode, the dominant current injection mechanisms can be identified for different materials and even for phase transitions within the material at the device-relevant length scales.

\section{Conclusions}
To conclude, we have demonstrated that by controlling the applied pressure  with an AFM tip, we can  non-invasively manipulate and control the electronic metal-insulator transition (MIT) of \ce{VO2}. 
Furthermore, we have shown that the metal-semiconductor junction plays an important role in the conduction mechanism of the nanometer-sized contact. 
Such mechanical tunability allows the experimental determination of the contact properties as well as the coupled structural and electronic metal-insulator transition (MIT). 
Specifically, the MIT in \ce{VO2} is observed in the barrier height parameter with varying temperature and pressure, respectively, in a hitherto inaccessible regime of the phase space. 
Using a pressure of approx. \SI{5}{\GPa}, we can reversibly and deterministically trigger the MIT at room temperature and mechanically switch the material locally, i.e., on the nanoscale under the probe tip, from an insulating to a conductive state.
At all times, an intrinsically insulating surface layer is inevitably present between the metal tip and the bulk-like interior of the \ce{VO2} thin film. 
Therefore, the dominating current injection mechanism is direct tunneling, which remains unaltered during the transition. 
The barrier height parameter changes significantly and abruptly during the metal-insulator transition. 
With varying temperature, only the band structure of the \ce{VO2} is changing, while with varying pressure, the thickness of the interfacial layer is affected in addition to the band structure. 
The combination of heteroepitaxial in-plane confinement due to fully coherent crystal growth of a thin layer onto a slightly mismatched substrate with local, uniaxial-like out-of-plane strain due to nano-indentation with a scanning probe tip allowed us to enter a hitherto inaccessible and unexplored region in the pressure-temperature phase diagram of \ce{VO2}, and serves as a versatile technique to study quantum materials under symmetry-breaking strains that can easily be extended towards very high electric fields and large strains, as well as spatial mapping of local differences in micro- and nanostructures. 
Our findings show that C-AFM is capable of measuring the contact properties of nanoscale junctions during phase transitions and to extract the dominant charge injection mechanisms as well as the barrier height parameters, which is valuable information for the development of future nanoscale metal-oxide devices.

\section{Experimental Section} 
\subsection{Sample preparation and structural characterization}
The thin films are deposited via reflection high-energy electron diffraction (RHEED)-assisted pulsed laser deposition (PLD) onto \SI{0.5}{\wtpercent} \ce{Nb}-doped \ce{TiO2} (001) substrates purchased from CrysTec GmbH. 
X-ray diffraction reciprocal space mapping is used to ascertain the coherent epitaxial relation between film and substrate and to extract the lattice parameters. 
Details of the sample preparation as well as the XPS/HAXPES and STEM-EELS analysis are reported in the Supp. Mater.

\subsection{AFM measurements}
All the C-AFM measurement are performed using a Bruker Dimension Icon AFM in a \ce{N2} environment by continuously purging with \ce{N2} gas, to avoid discrepancies in the data and water-induced reactions \cite{Schrecongost2019}. 
The relative humidity (RH) was measured using a humidity sensor (SENSIRION EK-H4 SHTXX, Humidity Sensors, Eval Kit, SENSIRION, Switzerland), with an accuracy of \SI{1.8}{\percent} between \num{10} and \SI{90}{\percent} RH. With the DCUBE mode, both current-voltage (\SI{}{\curr}(\SI{}{\volt})) and force-distance ($F(D)$) curves are obtained and form a hyperspectral dataset. Within this mode, the force is constantly regulated and recorded. The voltage is applied to the sample while the tip is grounded. The measurements are performed with highly boron-doped diamond (BDD) tips (AD-2.8-AS and FM-LC; Adama Innovations Ltd., resistivity: \SIrange[]{0.003}{0.005}{\ohm.\cm}).  The sample is heated using a platinum resistive type heater in a ceramic body and a tungsten cap controlled by a thermal applications controller (TAC). The samples are glued with silver paint on a metallic plate which is magnetically mounted onto the heater. A thermocouple is also mounted on the metallic plate to determine the temperature of the sample.
In order to determine the pressure with which the AFM tip is pressing on the sample surface, the contact area is needed. The contact area of the tip is acquired using the DMT theory \cite{Derjaguin1975}. For more detailed information see Supp. Mater. The error analysis for the measurements in Figure 1d and 2b can also be found in the Supp. Mater.

\section*{Conflict of interest}
The authors declare no competing interests.

\section*{Availability of data and materials}
The datasets generated during the current study are available from the corresponding author on reasonable request.

\section*{Author contributions}
Y.A.B and K.S. contributed equally.
Y.A.B. and K.S. conceived the project under supervision of G.K. and G.R.
Y.A.B. fabricated the samples and designed the simulations, performed the XRD and macroscopic transport experiments.
E.M., Y.B. and Y.A.B. performed and analyzed the XPS measurements.
K.S. performed and analyzed the AFM measurements.
K.S. and H.J.W.Z. interpreted the AFM data.
N.G., L.R., and D.J. collected and analyzed the STEM-EELS data under guidance of J.V.
Y.A.B. and K.S. wrote the manuscript with input from all authors.

\section*{Supporting Material}
Supporting Material is available from the Wiley Online Library.

\section*{Acknowledgements}
The authors would like to thank Dr. P. Lucke for fruitful discussions, and Y. Smirnov and M. Smithers for technical assistance.

\section*{Funding statement}
This work received financial support from the project Green ICT (grant number 400.17.607) of the research program NWA which is financed by the Dutch Research Council (NWO), Research Foundation Flanders (FWO grant number G0F1320N), and the European Union's Horizon 2020 research and innovation program  within a contract for Integrating Activities for Advanced Communities (grant number 823717 – ESTEEM3).
The K2 camera was funded through the Research Foundation Flanders (FWO-Hercules grant number G0H4316N - ``Direct electron detector for soft matter TEM'').


\clearpage
\bibliography{bibliography.bib}




\end{document}


\clearpage
\tableofcontents
\clearpage

\section{Sample preparation}	
\begin{figure}[tbh]
	\centering
	\includegraphics[width=85mm]{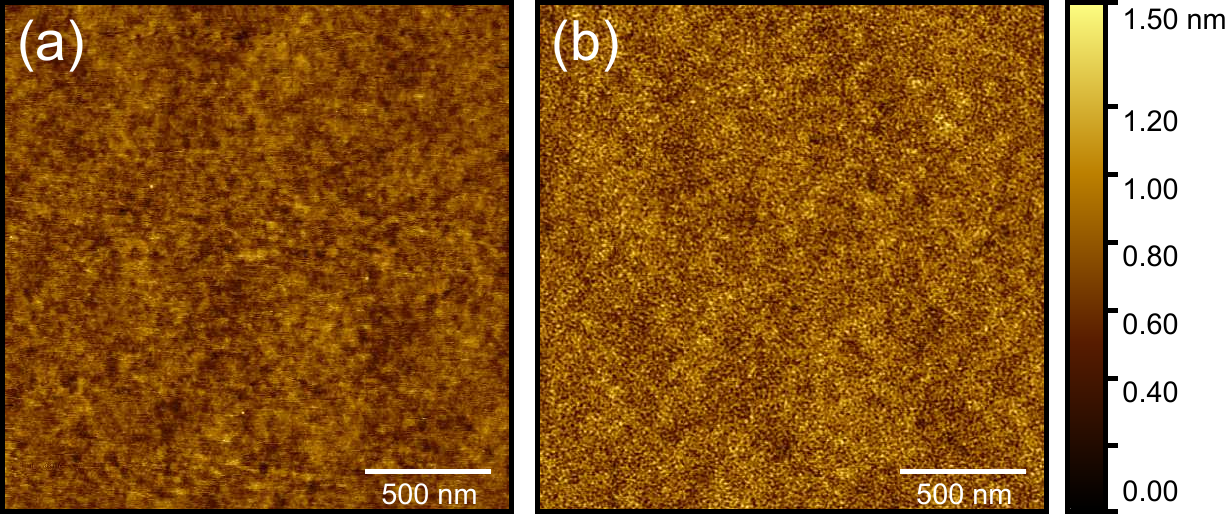}
	\caption{AFM Topography images (a) before (rms roughness \SI{0.13}{\nano\meter}) and (b) after (rms roughness \SI{0.18}{\nano\meter}) \ce{VO2} film growth.}
	\label{Fig_AFM_Topography}
\end{figure}

Single crystalline substrates of \SI{0.5}{\wtpercent} \ce{Nb}-doped rutile \ce{TiO2} (001) (5 x 10 x \SI{0.5}{\milli\meter}, miscut < \SI{0.2}{\degree}) were purchased from CrysTec GmbH. Prior to thin film growth, the substrates were cleaned in an ultrasonic bath for 5 minutes first with acetone and second with isopropanol. Subsequently, they were annealed for 120 minutes at \SI{800}{\celsius} in a tube furnace under an oxygen flow of \SI{150}{\liter\per\hour}. 
The resulting surface had a root mean square (rms) roughness of \SI{0.13}{\nano\meter} as determined by ambient tapping mode AFM (see Fig. \ref{Fig_AFM_Topography}) before loading the sample into the pulsed laser deposition (PLD) tool. Preliminary attempts to form an atomically flat surface with vicinal steps and terraces by annealing at \SI{850}{\celsius} (after 1 minute of etching in buffered HF) or \SI{950}{\celsius} for 90 minutes lead to complete destruction of the smoothly polished surface and excessive faceting.
 
The PLD used has a base pressure better than \SI{2E-7}{\milli\bar}.
Using a rectangular mask (3 x \SI{8}{\milli\meter\squared}), a mirror and a lens, \SI{248}{\nano\meter} wavelength UV light from a KrF excimer laser (COHERENT COMPex Pro 205 F, pulse duration \SI{25}{\nano\second}) was sharply imaged onto a raster scanning polycrystalline \ce{V2O5} target (\SI{20}{\milli\meter} diameter) with a repetition rate of \SI{4}{\hertz}. The spot size on the target was \SI{1.75}{\milli\meter\squared}. A manual attenuator in the beam path was used to adjust the laser fluence to \SI{1.3}{\joule\per\centi\meter\squared}. Before loading the substrate into the chamber, the target (ground with 1000 grit sandpaper) was pre-ablated with 2000 laser pulses.
The films were grown at a temperature of \SI{400}{\celsius} (measured by a thermocouple inside the resistive heater element) and under a dynamic background pressure of \SI{0.01}{\milli\bar} oxygen gas. This rather low temperature was chosen with the intention to create a sharp interface and limit titanium diffusion into the film.
After deposition, the sample was cooled down to room temperature at a rate of \SI{10}{\celsius\per\minute} under the deposition pressure. No additional annealing step was performed. 

\begin{figure}[tbh]
	\centering
	\includegraphics[width=140mm]{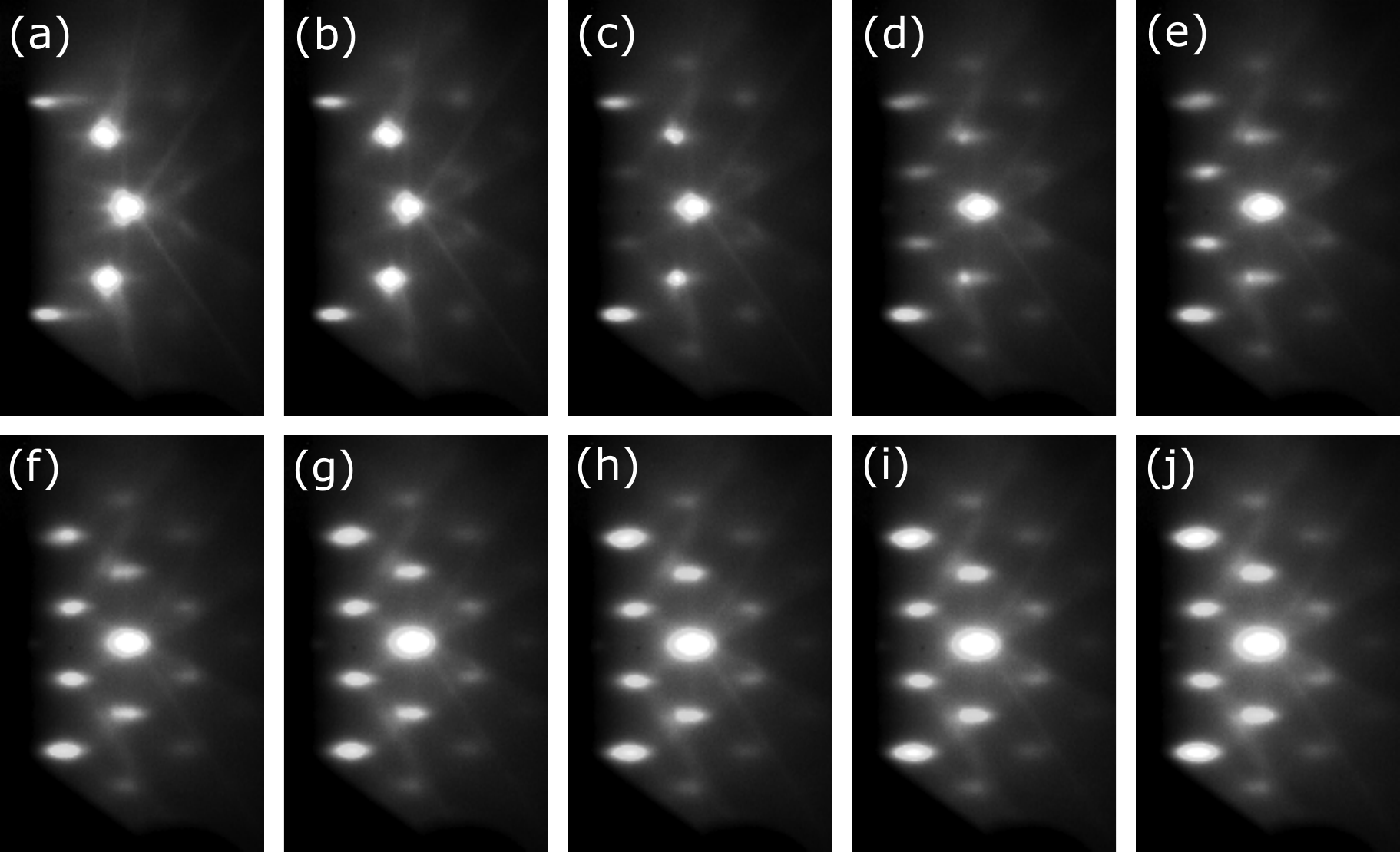}
	\caption{RHEED patterns taken before (a) and during deposition at approx. (b) 2, (c) 3, (d) 4, (e) 5, (f) 7, (g) 20, (h) 30, (i) 40, (j) 53 (end of the deposition) unit cells of \ce{VO2} film growth.}
	\label{Fig_RHEED}
\end{figure}

The PLD is equipped with \textit{in situ} high-pressure reflection high-energy electron diffraction (RHEED). In Fig. \ref{Fig_RHEED}, we show the diffractograms recorded at \SI{30}{\kilo\volt} in \SI{0.01}{\milli\bar} \ce{O2} along the [100] direction of the (001) rutile substrate before and during \ce{VO2} film growth. The sharp spots and the Kikuchi bands attest to a high surface quality. We monitored the intensity of the specular spot during deposition (not shown) but did not observe any intensity oscillations that would be expected from layer-by-layer growth. The observed diffractogram after film growth is a typical transmission pattern characteristic of an island growth mode, consistent with the absence of intensity oscillations in the specular spot. It is likely that the unfavorably high surface energy of the (001) plane is the underlying cause for promoting the island growth mode and associated surface roughening \cite{Krisponeit2020, Wahila2020}.

\clearpage	
\section{Structural characterization}
High-resolution X-ray diffraction (XRD) was performed using a Bruker D8 Discover equipped with a high brilliance microfocus rotating anode generator (TXS, \SI{2.5}{\kilo\watt}), an asymmetric channel-cut Ge (022) two-bounce monochromator, a \SI{1}{\milli\meter} pinhole collimator, and a large Eiger2 R 500k area detector with high dynamic range.

\begin{figure}[tbh]
	\centering
	\includegraphics[width=150mm]{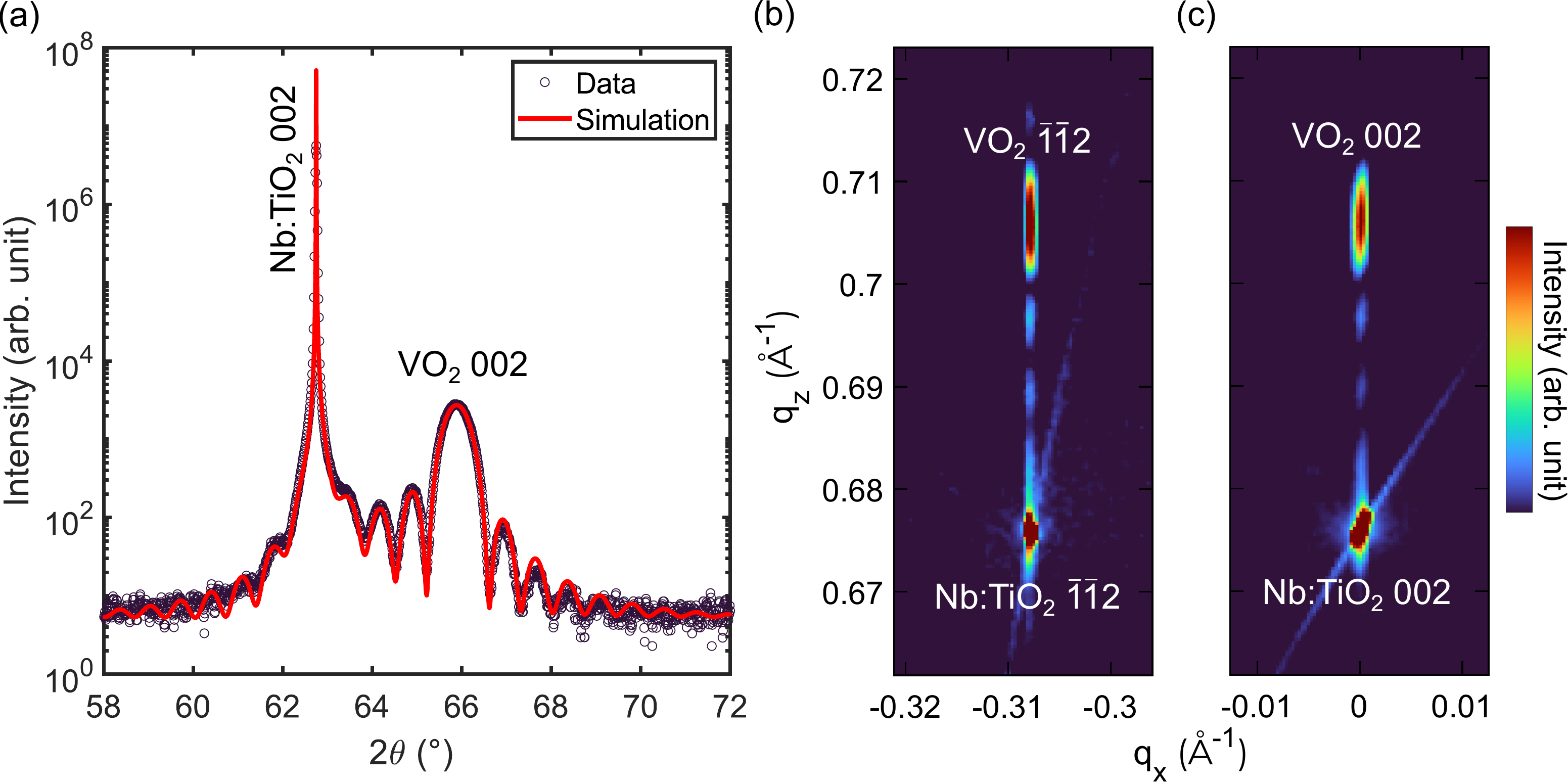}
	\caption{X-Ray diffraction. 
		(a) $2\theta - \omega$ scan with a superimposed fit. 
		(b) Asymmetrical and (c) symmetrical reciprocal space maps  of the $\bar{1} \bar{1}2$ and $002$ reflections confirming the single-oriented growth and coherently strained epitaxy as evidenced by the identical in-plane ($q_x$) momentum transfers for substrate and thin film.}
	\label{Fig_XRD}
\end{figure}

The \ce{VO2} film thickness was determined by fitting the Laue fringes observed in a $2\theta - \omega$ scan (see Fig. \ref{Fig_XRD} (a)) via a dynamical X-ray diffraction simulation using the program \textit{gid\_sl} on Sergey Stepanov's X-ray server (\url{https://x-server.gmca.aps.anl.gov}) \cite{Stepanov2004}. In the calculation, we assume an interface roughness of \SI{0.2}{\nano\meter} consistent with the AFM measurements shown in Fig. \ref{Fig_AFM_Topography}. It is those sharp interfaces combined with the high crystallinity that enable the observation of multiple Laue fringes.
For a deposition time of 1000 seconds with 4000 laser pulses at \SI{4}{\hertz} a film thickness of \SI{15.0 \pm 0.1}{\nano\meter} is extracted, resulting in a growth rate of \SI{0.015}{\nano\meter\per\second} which corresponds to \SI{75(1)} pulses per unit cell.

Three-dimensional reciprocal space maps (RSM) were constructed from sets of rocking curves measured in coplanar geometry. Projections summed over one of the in-plane momentum axes are shown in Fig \ref{Fig_XRD} (b - c).
From these RSMs the following rutile-like lattice parameters for the \ce{VO2} film are extracted via 2D Voigt function fitting of the thin film peak:
$a = b = $ \SI{4.5937(10)}{\angstrom}, $c = $ \SI{2.8275(50)}{\angstrom}.

As films with a thickness of \SI{15}{\nm} or more are prone to cracking over time to relax the strain, the remainder of this study, in particular the AFM, XPS/HAXPES, and STEM-EELS experiments, was performed on thinner films with a thickness of only \SI{10}{\nm}.

\clearpage
\section{Tip and sample parameters}
\begin{figure}[tbh]
	\centering
	\includegraphics[width=165mm]{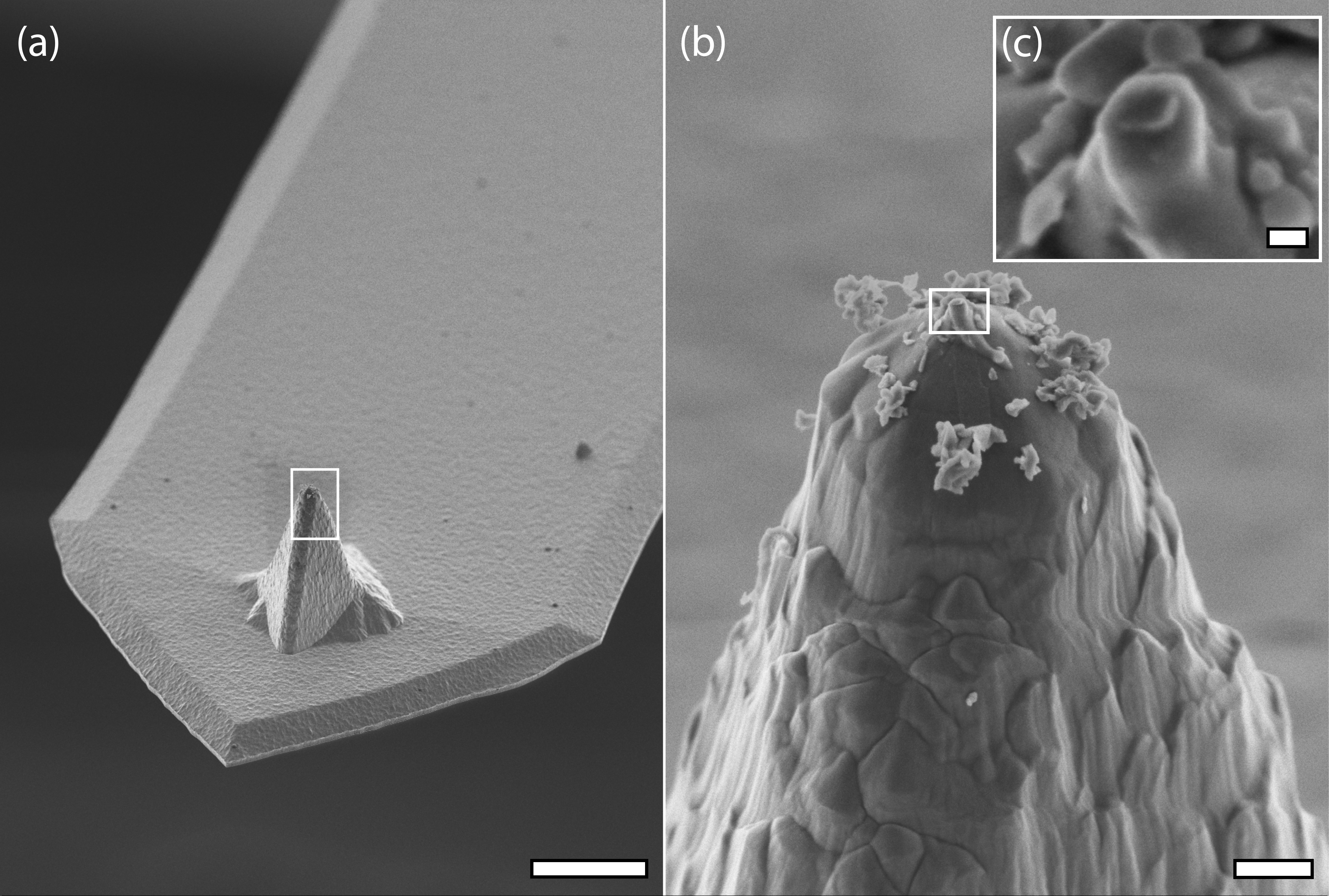}
	\caption{Scanning electron micrographs of the tip used for this study (AD-2.8-AS). Panel (b) is a zoom of the area indicated by the white rectangle in (a). Panel (c) is an even higher magnification of the area indicated by the white rectangle in (b), albeit recorded at a slightly different angle.
		The scale bars in panels (a), (b), and (c) measure \SI{10}{\um}, \SI{1}{\um}, and \SI{50}{\nm}, respectively.}
	\label{Fig_SEM}
\end{figure}

In Table \ref{TipTable} and \ref{SampleTable}, the material properties of the boron-doped diamond tips (purchased from Adama Innovations Ltd., Ireland) and the sample are summarized. The work function of the diamond tip in Table \ref{TipTable} is characterized using Kelvin Probe Microscopy. The radius is confirmed by SEM measurements to be within the range specified by the supplier, namely $r = $\SI{10(5)}{\nm} (see Fig. \ref{Fig_SEM}). The spring constant is determined by using the thermal vibration method \cite{Hutter1993}. First, the deflection sensitivity is determined on a sapphire sample, and subsequently, the spring constant is measured from the thermal tuning.

By measuring the I(V) characteristics of a Au film (not shown) and confirming that it is ohmic as expected, we verified that the tunneling behavior shown in the main article is indeed a property of the \ce{VO2} - tip junction, and not a property inherent to the boron-doped diamond tip alone.

\begin{center}
\begin{table}
\begin{tabular}{|l || c | c | c| c| c|c|c| } 
 \hline
 {} & $Y$ [GPa] & $\nu$ & $k$ [N/m] & $f_\text{r}$ [kHz] & $\phi_\text{M}$ [eV] & $r$ [nm] & $\chi$ [eV] \\ 
  \hline\hline
 AD-2.8-AS & 1000 \cite{Klein1992}  &  0.07 \cite{Klein1992} & 8 & 65 & 5.1  & 15 & 0.02 \cite{Liu1997} \\ 
 \hline
FM-LC & 1000 \cite{Klein1992}  &  0.07 \cite{Klein1992} & 3 & 65 & 5.1  & 30 & 0.02 \cite{Liu1997} \\ 
\hline
\end{tabular}

\caption{Various cantilever properties including the Young's modulus ($Y$), Poisson ratio ($\nu$), spring constant ($k$), work function ($\phi_\text{M}$) and tip radius ($r$). The spring constant is determined using the thermal vibration method \cite{Hutter1993}.}
\label{TipTable}
\end{table}
\end{center}

\begin{center}
\begin{table}
\begin{tabular}{|c || c | c | c| c|} 
 \hline
 {} & $Y$ [GPa] & $\nu$ & $E_\text{G}$ [eV] & $\chi$ [eV] \\ 
  \hline\hline
 VO$_2$ & 90-150 \cite{Guo2011,Fan2009,Sepulveda2008,Tsai2004} & 0.249 \cite{Aetukuri2013} & 0.6-0.7\cite{Lee2015, Yoon2016} & 2 \cite{Zhang2020}\\ 
 \hline
 TiO$_2$ & 151 \cite{Borgese2012} & 0.27\cite{Borgese2012} & 3.1 \cite{Pascual1978} &  3.43 \cite{SinghSurah2019} \\
 \hline
\end{tabular}

\caption{Various bulk material properties including the Young's modulus ($Y$), Poisson ratio ($\nu$), band gap ($E_\text{G}$) and electron affinity ($\chi$).}
\label{SampleTable}
\end{table}
\end{center}

\clearpage
\section{Finite element model}
\begin{figure}[tbh]
	\centering
	\includegraphics[width=\columnwidth]{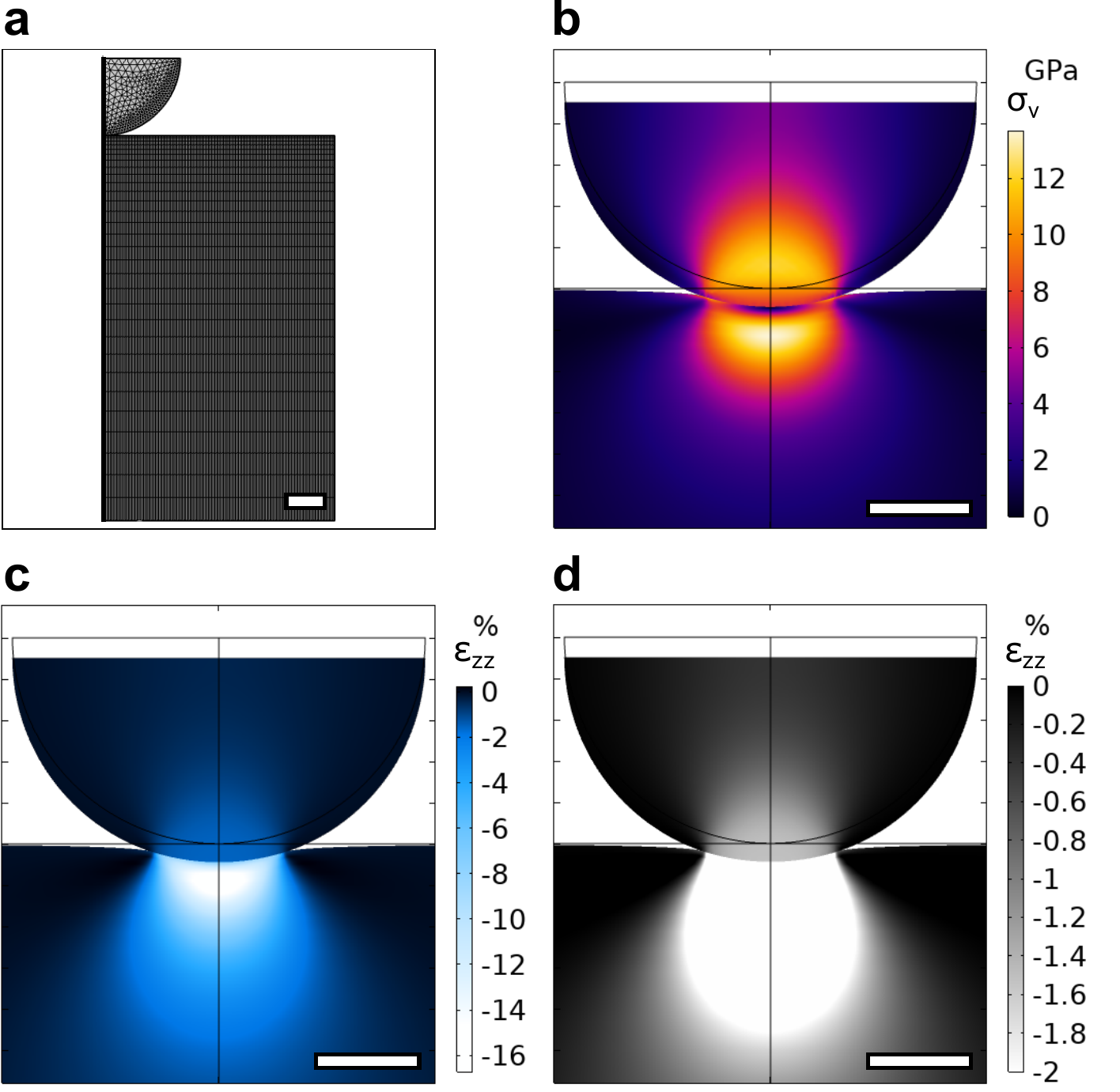}
	\caption{Finite element model. 
		(a) Drawing of the mesh,
		(b) von Mises stress $\sigma_v$ distribution in the fully indented state, where the tip is displaced by \SI{1}{\nano\meter},
		(c) out-of-plane compressive strain $\epsilon_{zz}$,
		(d) same data as in (c) but on truncated greyscale to emphasize the part of the film which experiences at least \SI{2}{\percent} of compressive strain, which is expected to induce the phase transition. The position and shape of the indenter as well as the film prior to the forced displacement are sketched with thin black lines in (b) - (d).
		The length of the scale bar in all panels is \SI{5}{\nano\meter}.}
	\label{Fig_FEM}
\end{figure}

A finite element model of the nanoindentation process is made in COMSOL Multiphysics 5.6 using the \textit{MEMS Module}.  The indentation process is implemented via a parametric deformation study. The apex of the AFM probe is modeled as a half-sphere, which is indented into the stationary \ce{VO2} layer. The bottom of the \ce{VO2} layer is constrained by fixed boundary conditions, whereas the tip of the indenter undergoes a prescribed parametric displacement downwards into the film. The sides are not constrained in the simulation. 2D axisymmetry along a vertical line is applied to the simplified cross-sectional geometry, see the drawing in Fig. \ref{Fig_FEM} (a).
As usual, the mesh is refined near the contact line and finer for the impacted surface than for the indenter.
Triangular elements are used for the indenter and rectangular elements for the film.
For simplicity, the Nb-doped \ce{TiO2} substrate is omitted, but to nevertheless assure a homogeneous stress and strain distribution far away from the indenter, the film is modeled much thicker in the finite element simulation than in the actual experiment described in the main article. 
The materials parameters used in the simulation are the same as summarized in the tables in the previous section of this supplemental material. The simulation does not include the effects of adhesion as these were beyond the scope of our experimental investigation.

The stress and strain distributions resulting from an indentation of \SI{1}{\nano\meter} are depicted in Fig. \ref{Fig_FEM} (b) - (d).
The peak compressive out-of-plane strain $\epsilon_{zz}$ exceeds \SI{16}{\percent} locally under the tip, but more importantly for this study, the entirety of a $\sim$ \SI{10}{\nano\meter} thick layer experiences a compressive out-of-plane strain of more than \SI{2}{\percent}, which we expect to suffice to induce the phase transition in \ce{VO2} at room temperature based on an extrapolation of the published pressure-temperature phase diagram by \citet{Park2013}.
Not shown here are the in-plane strains and stresses, which are much smaller than the relevant out-of-plane counterparts.

We would like to note that more than half of the strain needed to trigger the phase transition in \ce{VO2} at room temperature (\SI{2}{\percent} compression) is already statically and non-reversibly provided by epitaxy, as the \ce{VO2} thin film is commensurately strained onto the Nb-doped \ce{TiO2} (001) substrate. As shown by our XRD measurements, the lattice mismatch leads to a contraction of the rutile c-axis of the \ce{VO2} thin film of $\sim$ \SI{1.2}{\percent}. Hence, much less than the \SI{1}{\nano\meter} indentation simulated in Fig. \ref{Fig_FEM} should suffice to demonstrate the pressure-driven semiconductor-metal transition.

Due to the gross simplification of the geometry and some uncertainty in the materials parameters, the simulation should be taken only as a ballpark estimate and not as a quantitatively accurate prediction. We would like to note, however, that the order of magnitude of the calculated stresses and strains are in reasonable agreement with the experimental findings presented in the main article.

\clearpage
\section{Acquisition of the $I(V)$ curves} \label{IVacquire}
All the measurements in this study are obtained using the Bruker DCUBE mode. Within this mode, both the force-distance curves ($F(D)$) and current-voltage ($I(V)$) curves are obtained. During the $I(V)$ measurement, the force is actively controlled and monitored. This avoids any change in force during the $I(V)$ measurement. Besides the control during the electrical measurement, also the force during the approach and the retraction is measured. From these force-distance curves, additional information can be extracted, such as the adhesion and the amount of indentation \cite{Butt2005}. However, the most important parameter extracted from these measurements is the total work of adhesion which is essential for the determination of the contact area (see section \ref{ContactArea}) \cite{Carpick1999}.

\begin{figure}[tb]
	\centering
	\includegraphics[width=85mm]{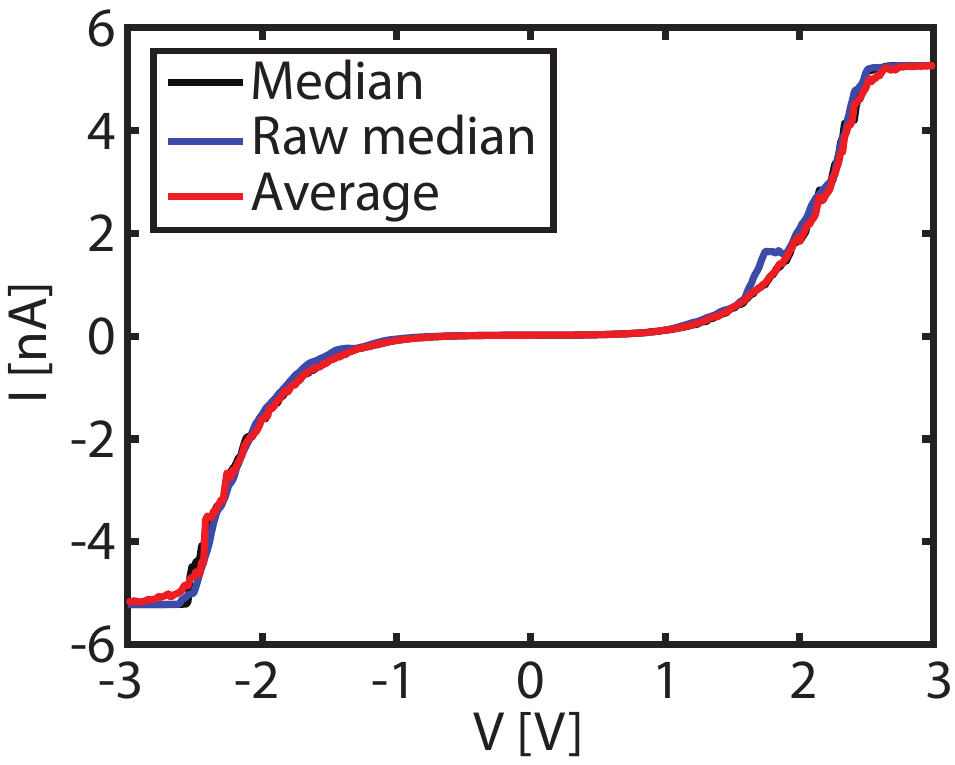}
	\caption{An $I(V)$ curve recorded on VO$_2$ at 50 $^\circ$C and 4.2 GPa. The median curve is used for the data analysis. The raw median curve is too noisy to be used in the analysis.}.
\label{Fig_Median_curve}
\end{figure}

The $I(V)$ curves shown in the main text are obtained by taking the median over multiple curves. Multiple $I(V)$ curves are taken in a grid-like fashion, in which the scanning area is divided in 12 $\times$ 12 (or 8 $\times$ 8) equally spaced positions. At every position, an $I(V)$ measurement is recorded. The average spacing between the spectra is approx. 1 $\mu$m. For each $I(V)$ curve the piezo scanner is paused, and the bias is ramped from -3 V to 3 V to avoid permanent deformation of the sample \cite{Schrecongost2019}. From every measurement, the average curve, the median curve, and the raw median curve are extracted from 144 curves in total. The average curve is the mean of all the 144 curves, while the median curve is the median of the same measurements. The raw median curve is determined by determining the median of every single point of the $I(V)$ curve (which is 416 points). The most common curve number is then taken as the raw median curve. As can be seen in Fig. \ref{Fig_Median_curve}, there is no deviation between the median curve and the raw median curve. As the median curve is much smoother, this is the curve used for the analysis in the main text. The average curve deviates from most of the measurements around the saturation point of the I-V converter ($I > 5$ nA in this example) and is therefore not used.

\section{Stability of the \ce{VO2} layer}
After the pressure experiments, part of the surface where the measurements are performed are investigated with an AFM. No permanent deformation or damage is observed to the layer, proving the reversibility of the metal-insulator transition. However, when the surface is scanned with the same force as the $I(V)$ measurements are performed, the VO$_2$ layer is scraped away from the TiO$_2$ substrate. The lateral movement of the tip inflicts damage on the surface and then acts as a scalpel. Therefore, it is crucial to only exert pressure in the vertical direction on the surface and avoid any lateral movement.

\begin{figure}[tb]
	\centering
	\includegraphics[width=150mm]{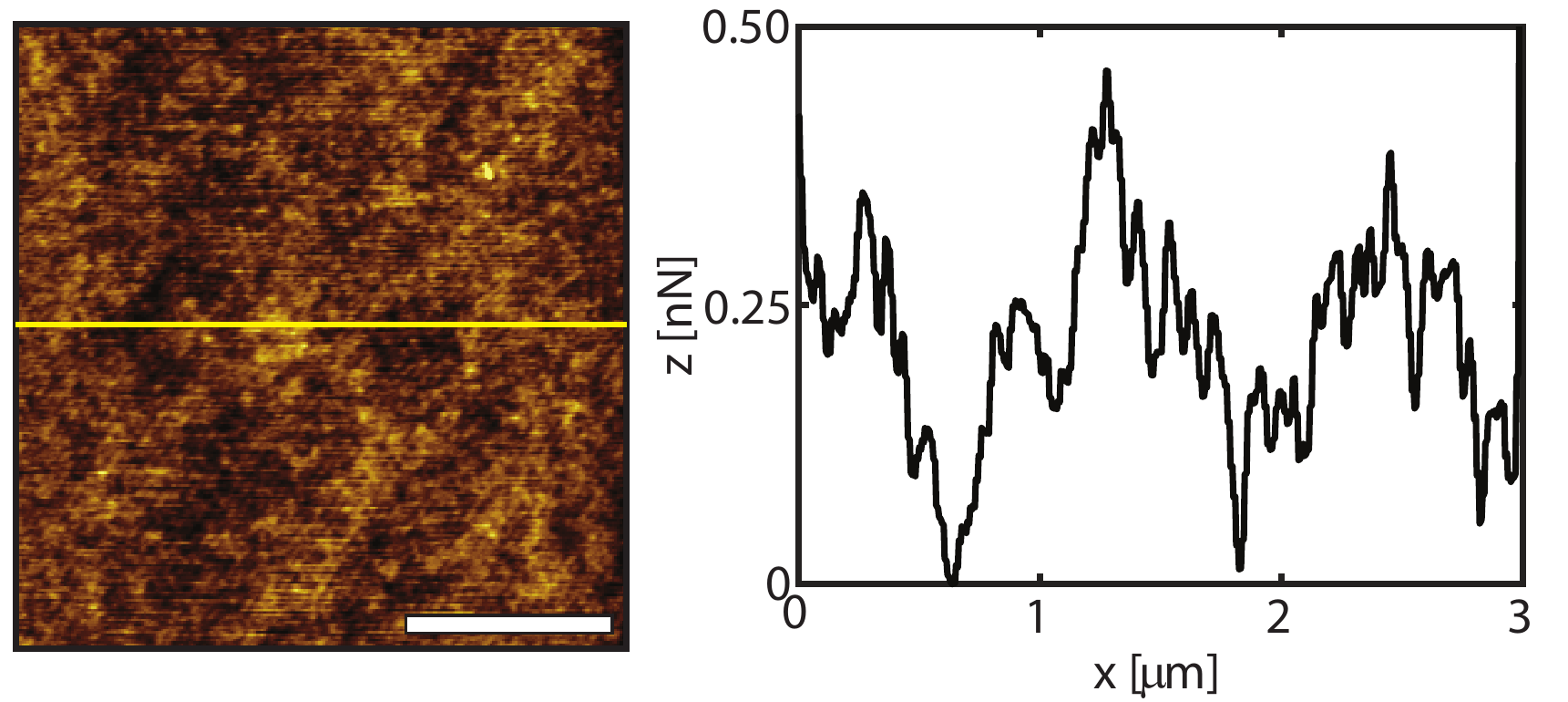}
	\caption{(a) Topographic image (3 $\times$ 3 $\mu$m, scale bar 1 $\mu$m) of the VO$_2$ surface after a pressure experiment. No permanent deformation or damage is observed after the experiment. (b) Cross-section height profile of the yellow line in (a).}.
\label{Fig_Control_Image}
\end{figure}

\section{Contact area} \label{ContactArea}
In order to determine the applied pressures with which the AFM cantilever presses on the host material, the applied load, and the effective contact area need to be known. The applied load is controlled by the measurement sequence described in section \ref{IVacquire}. The effective contact area can be derived from contact-mechanic models such as the Hertzian model, which assumes a hemispherical tip shape and neglects adhesive forces \cite{Hertz1881}.

Based on the Hertz equation, the physical contact area radius of the tip on a surface is given by

\begin{equation}
a = \sqrt[3]{RFK}
\label{Hertz}
\end{equation} 
\\ where $R$ is the radius of the tip, $F$ the applied load, and $K$ is the combined elastic modulus of the tip and sample, given by $K = 4/3((1-\nu_\text{t}^2)/Y_\text{t})^{-1}+((1-\nu_\text{s}^2)/Y_\text{s}))^{-1}$ where $Y_\text{t}$, $\nu_\text{t}$ $Y_\text{s}$ and $\nu_\text{s}$ are the Poisson ratios and Young's moduli of tip and sample, respectively.

In the Hertz model, the adhesion of the sample is neglected, whereas two other theories do take the adhesion into account, the Johnson-Kendall-Roberts (JKR) \cite{Johnson1971} and Derjaguin-M{\"u}ller-Toporov (DMT) \cite{Derjaguin1975} theories. The JKR theory can be applied in the case of large tips and soft samples with a large adhesion, the DMT theory in the case of small tips and stiff samples with a small adhesion. The used tip has an approximated radius of $\approx$ \SI{20}{\nano\meter} and \ce{VO2} is a stiff sample (see table \ref{SampleTable}). Therefore, the DMT theory is used \cite{Carpick1999,Butt2005}.

\begin{figure}[th]
	\centering
	\includegraphics[width=85mm]{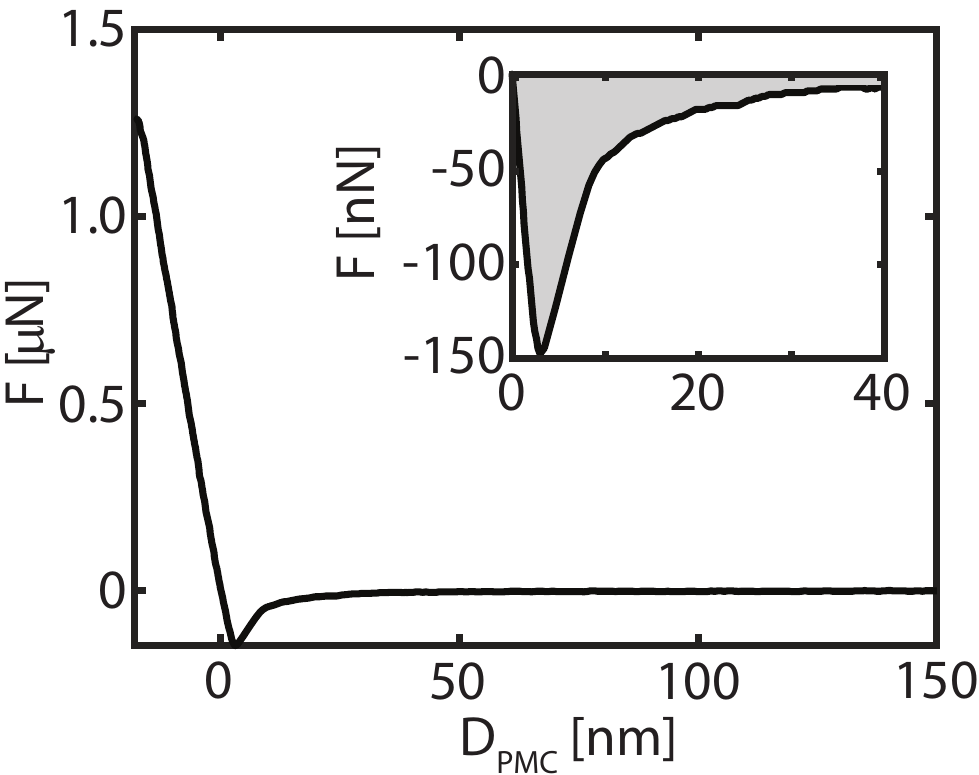}
	\caption{Force-distance curve of a VO$_2$ film at 50 $^\circ$C probed with a boron-doped diamond tip with a tuned $k$ = \SI{8}{\newton\per\meter}. The inset shows a zoom of the region near the origin. The gray marked region is defined as the work of adhesion per unit area, $W$.}
\label{AdhesionCurve}
\end{figure}

\clearpage
In the DMT theory, equation \ref{Hertz} is expanded to \cite{Derjaguin1975}

\begin{equation}
a = \sqrt[3]{RK(F+2\pi RW)}
\label{DMT}
\end{equation} 
\\ with $W$ the work of adhesion per unit area. As the tip is approaching the surface, the two surfaces are attracted towards another. This introduces an additional force besides the force with which the tip presses on the surface (which is chosen by the user). An example of an $F(D_\text{PMC})$ curve is shown in Fig. \ref{AdhesionCurve}, in which $D_\text{PMC}$ is the distance determined by the piezo motor controller. In order to get the displacement $D$, the information of the height sensor is also needed. The marked gray area in Fig. \ref{AdhesionCurve} is $W$ \cite{Carpick1999}. The slope of the curve is constant over all the extension of the piezo motor controller, indicating a pure elastic, non-indentation model for the tip-sample contact \cite{Domingo2015}.

\clearpage
\section{Charge injection mechanisms}

\begin{figure}[tb]
	\centering
	\includegraphics[width=150mm]{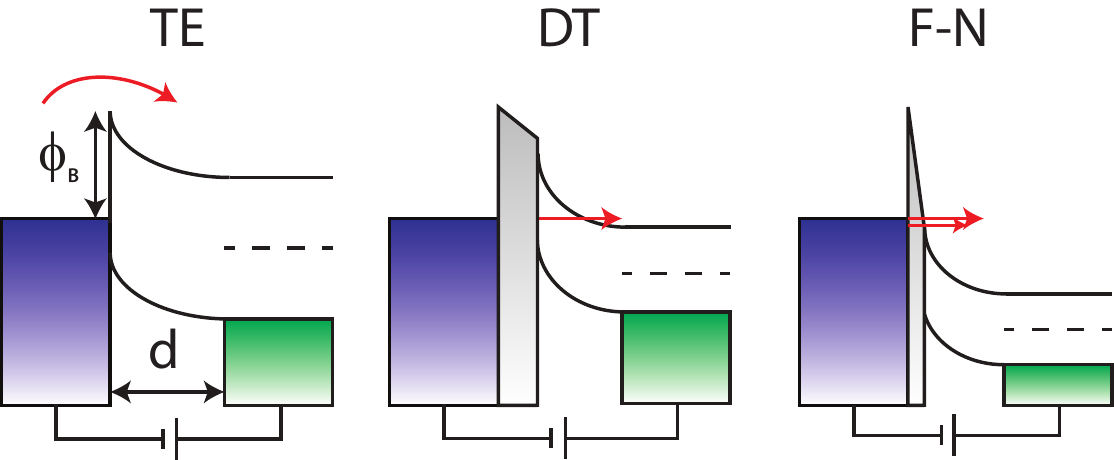}
	\caption{Schematic representation of the three injection mechanisms considered within the manuscript; thermionic emission (TE), direct tunneling (DT), and Fowler-Nordheim tunneling (F-N).}
\label{Fig_Models}
\end{figure}

From the measured $I(V)$ curves it is possible to acquire information about the Schottky barrier height ($\phi_\textrm{B}$) and the dominant charge injection mechanism. The three different charge injection mechanisms are thermionic emission (TE), direct tunneling (DT) and Fowler-Nordheim tunneling (F-N). For TE, the Schottky barrier height and the ideality factor ($\eta$, this is the deviation of the charge transport from ideal thermionic emission) can be extracted. Within the TE model, the emission current $I$ is given by \cite{Roderick1988,Sotthewes2019,Bremen2019} 

\begin{equation}
I = I_{0}\exp\Big(\frac{qV}{\eta k_\textrm{B} T}\Big) \bigg[1 - \exp\Big(\frac{-qV}{k_\text{B} T}\Big)\bigg]
\label{TE-eq}
\end{equation}
\\ where $V$ is the applied, $k_\textrm{B}$ is the Boltzmann constant, $T$ the temperature, and $q$ is the electron charge. For values of $V$ larger than $3k_\textrm{B}T/q$, the second term in equation \ref{TE-eq} becomes negligible, equation \ref{TE-eq} simplifies to:

\begin{equation}
I = I_{0}\exp\Big(\frac{qV}{\eta k_\textrm{B} T}\Big)
\end{equation}
\\ The saturation current  ($I_0$) is a constant and depends on $\phi_\textrm{B}$

\begin{equation}
I_{0} = AA^{*}T^{2}\exp\Big(-\frac{q \phi_\text{B}}{k_\text{B} T}\Big)
\label{SatCur}
\end{equation}
\\with $A$ the effective contact area between the AFM tip and the VO$_2$, calculated using the formulas in section \ref{ContactArea} and $A^{*}$ the Richardson constant ($A^{*} = \frac{4 \pi q m^* k_\textrm{B}^2}{h^3}$, with $m^*$ the effective mass and $h$ the Planck constant). The ideality factor can be obtained from 
\begin{equation}
\eta = \frac{q}{k_{\text{B}} T}\frac{d V}{d(\ln I)}
\end{equation}

and the Schottky barrier height ($\Phi_\textrm{B}$) is given by

\begin{equation}
\Phi_\text{B} = \frac{k_\text{B}T}{q} \ln\Big(\frac{A^*AT^2}{I_0}\Big)
\end{equation}
\\The ideality factor is extracted from the slope of the TE regime in the $\ln(I)-V$ plot, while the intercept of the curve gives the saturation current (see equation \ref{SatCur}) which is used to extract $\phi_\textrm{B}$.

From F-N and DT, a product of the Schotty barrier height and the barrier width, referred to as the barrier parameter, can be obtained. It should be noted that the barrier parameter of F-N differs from the barrier parameter of DT. In a F-N plot, the F-N regime can be recognized as it satisfies the linear relation \cite{Beebe2006,Ikuno2011,Bremen2019}

\begin{equation}
\ln\Big(\frac{I}{V^2}\Big) = \ln\Big(\frac{Aq^3m_0}{8\pi h \phi d^2 m^*}\Big)-\frac{8 \pi \sqrt{2m^*}\phi_\text{B}^{3/2}d}{3 h q V}
\end{equation}
\\ The slope of the F-N regime then equals $8 \pi \sqrt{2m^*}\phi^{3/2}d/3 h q V$, from which the barrier parameter $\phi^{3/2}d$ can be extracted.
The barrier parameter of DT can also be obtained from the F-N plot. The DT regime satisfies \cite{Beebe2006,Ikuno2011,Bremen2019}
\begin{equation}
\ln\Big(\frac{I}{V^2}\Big) = \ln\Big(\frac{Aq^2\sqrt{2 m^* \phi}}{Vh^2d}\Big)-\frac{4 \pi d \sqrt{2m^*\phi}}{h}
\label{DT_diff}
\end{equation}
\\ Within the F-N plot, this regime can be plotted with a logarithmic function. The right term in equation \ref{DT_diff} then equals $4 \pi d \sqrt{2m^*\phi}/h$, from which the barrier parameter $\sqrt{\phi}d$ can be extracted. Because F-N is absent within our measurements, the DT parameters in the main text are extracted from a $\ln(I/V^2)$ versus $\ln(|1/V|)$ plot.

\section{Error analysis}
\begin{figure}[th]
	\centering
	\includegraphics[width=150mm]{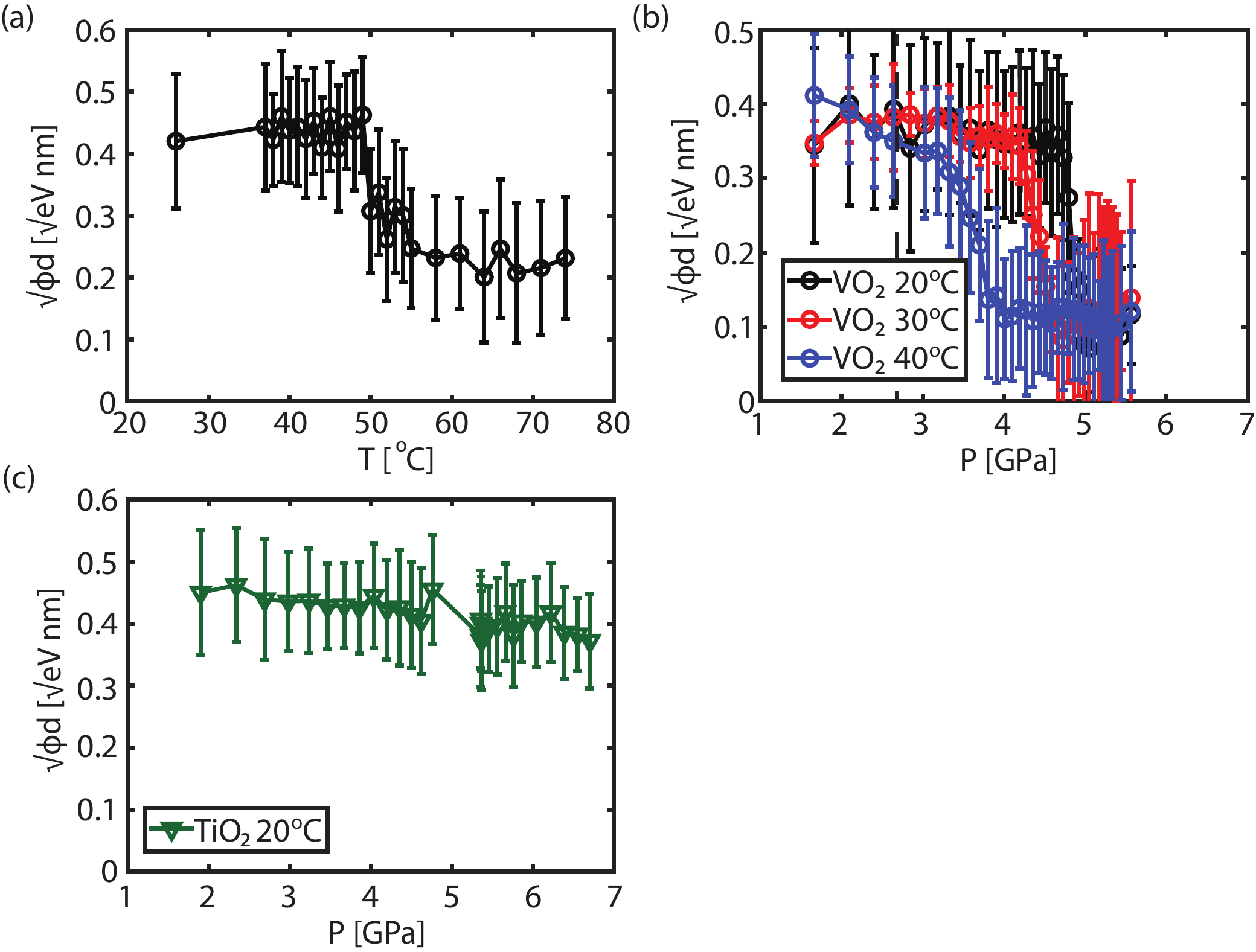}
	\caption{Figures 1D and 2B from the manuscript including error bars. (a) Fig. 1D of the manuscript. The obtained barrier height parameter versus the temperature for VO$_{2}$. (b) Fig. 2B of the manuscript. The obtained barrier height parameters versus pressure for VO$_2$ at different temperatures. (c) Fig. 2B of the manuscript. The barrier height parameter on TiO$_2$ versus pressure.}
\label{SuppFig_error}
\end{figure}

The uncertainty in the measurements is determined by calculating the standard deviation per measurement point in Fig. \ref{SuppFig_error}. Each measurement point in Fig. \ref{SuppFig_error} consists of 64-144 measurements (see section \ref{IVacquire}). From each individual curve, the barrier height parameter is extracted, and based on these values, the standard deviation is determined. The standard deviation is visualized as the error bars in Fig. \ref{SuppFig_error}. 

\section{Thermionic emission data}
\begin{figure}[th]
	\centering
	\includegraphics[width=150mm]{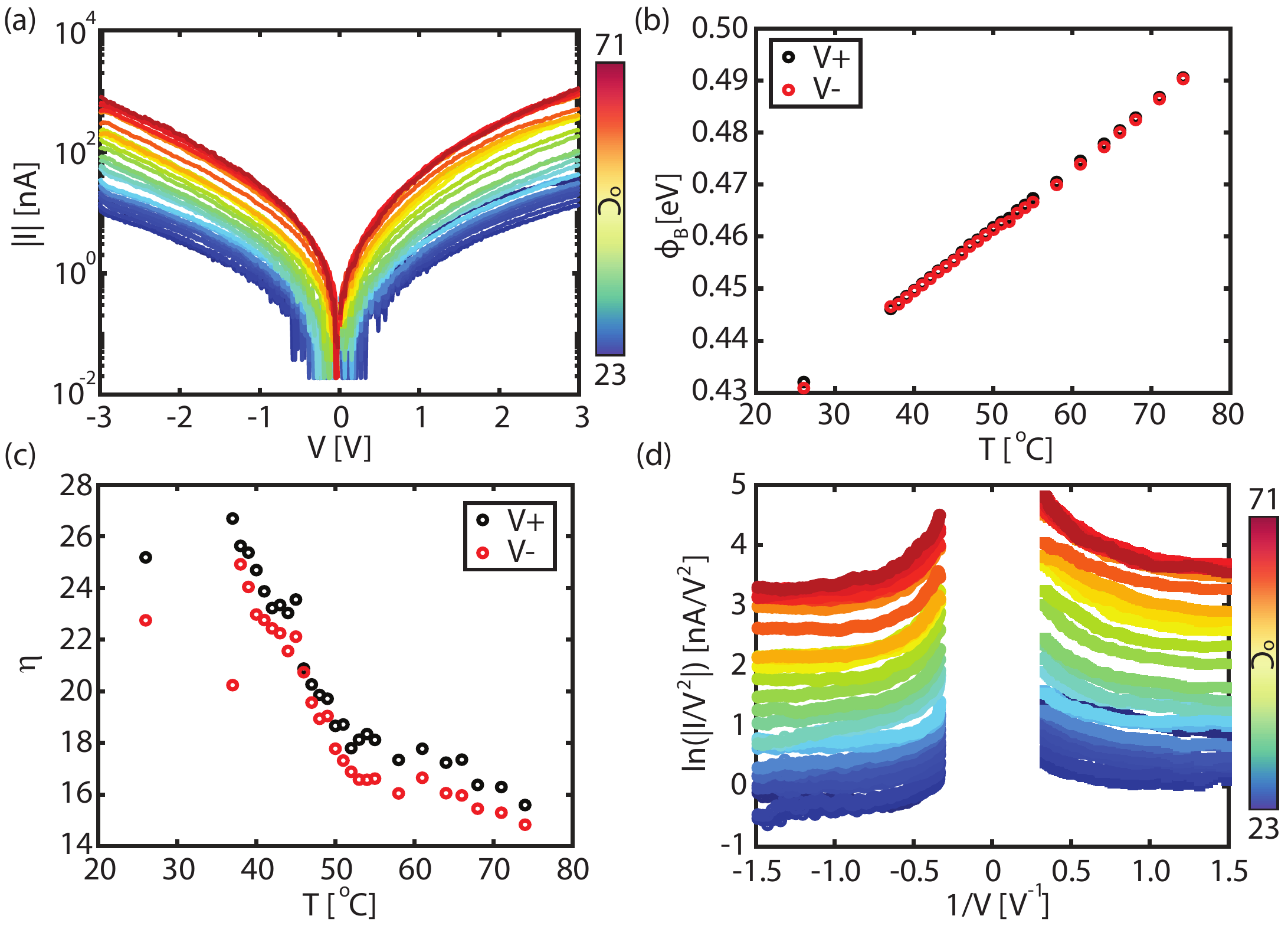}
	\caption{(a) The median $I(V)$ curves on \ce{VO2} on a semi-logarithmic scale for different temperatures. (b) The extracted Schottky barrier height versus the temperature extracted using the thermionic emission model. A linear increase in $\phi_\text{B}$ with temperature is observed. (c) The obtained ideality factor as a function of temperature. With increasing temperature, the ideality factor decreases because the electrons gain more energy to cross the barrier. (d) F-N plots of the data in (a). Almost no linear relation is observed related to F-N tunneling. Therefore, the dominant transport mechanism is DT.}
\label{TE_VarT}
\end{figure}

\begin{figure}[th]
	\centering
	\includegraphics[width=150mm]{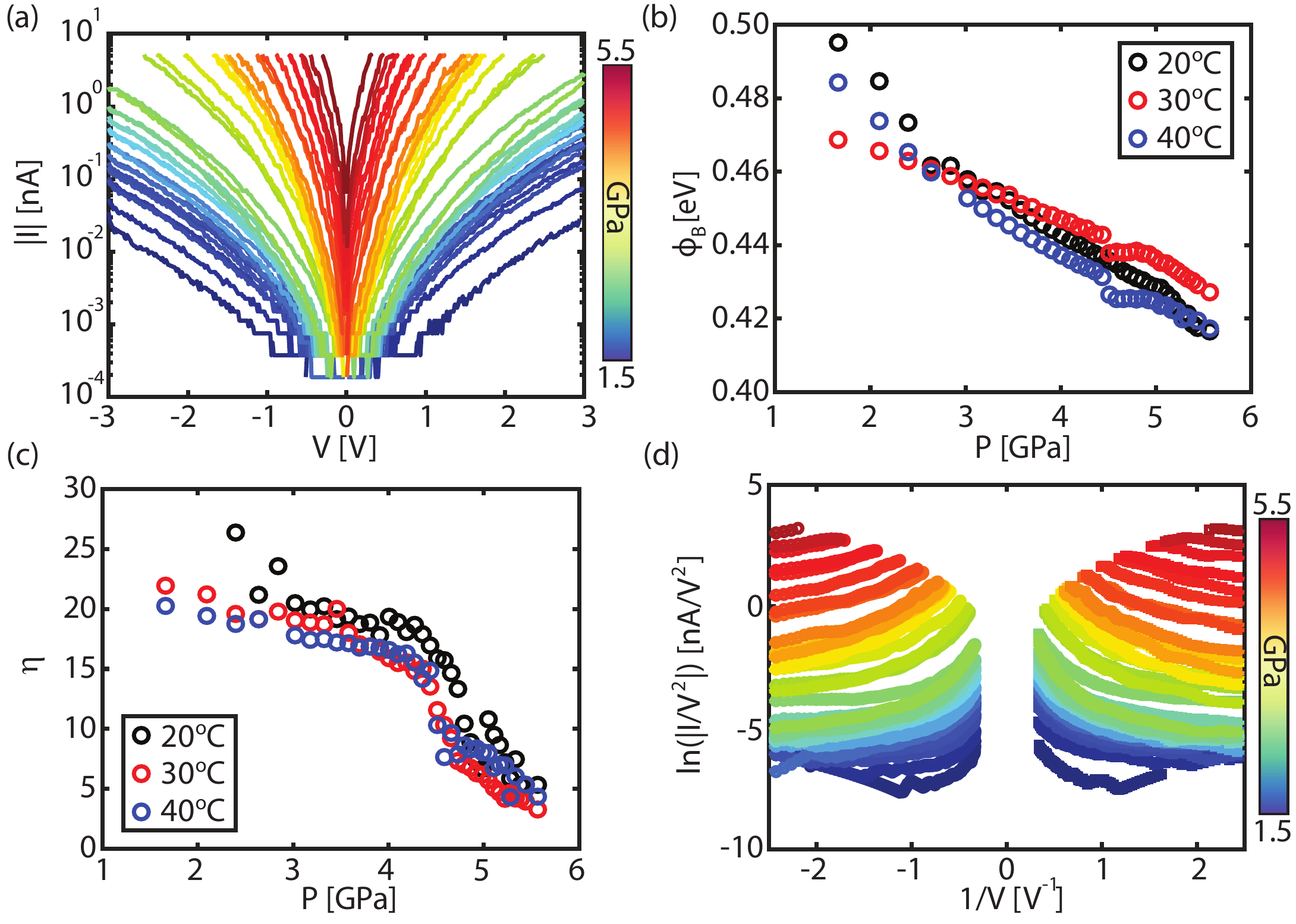}
	\caption{(a) The median $I(V)$ curves on \ce{VO2} on a semi-logarithmic scale for different pressures at a temperature of 50 $^\circ$C (b) The extracted Schottky barrier height as a function of pressure for three different temperatures. (c) The obtained ideality factor as a function of pressure for different temperatures. A change in slope is observed for all three temperatures at the same pressure. At higher pressures, the ideality factor drops to approx. 5. (d) (d) F-N plots of the data in (a). Almost no linear relation is observed related to F-N tunneling. Therefore, the dominant transport mechanism is DT.}.
\label{TE_VarP}
\end{figure}

\begin{figure}[th]
	\centering
	\includegraphics[width=150mm]{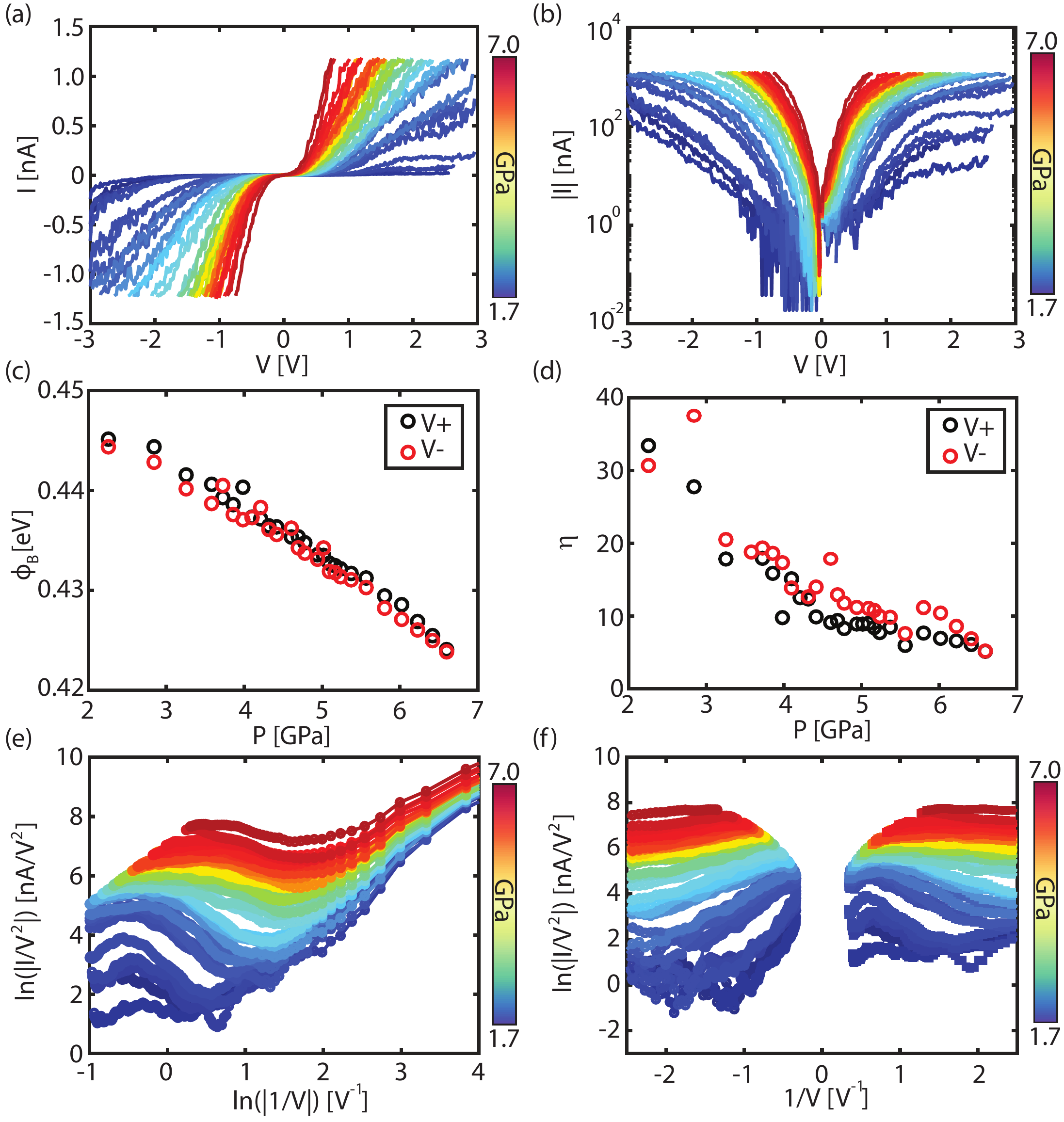}
	\caption{(a) The median $I(V)$ curves on TiO$_2$ for different pressures. (b) Same data as in (a) on a semi-logarithmic scale. (c-d) The extracted Schotty barrier height and ideality factor using the thermionic emission model (see equation \ref{TE-eq}) Similar to VO$_2$, thermionic emission is not the main charge injection mechanism ($\eta >$ 4). (e)  Same data as in (a) plotted on a $\ln(I/V^2)$ versus $\ln(|1/V|)$ scale (only the negative polarity is shown), which is related to the DT model. (f) F-N plots of the data in (a). Almost no linear relation is observed related to F-N tunneling. Therefore, the dominant transport mechanism is DT, very similar to VO$_2$.}
\label{TE_TiO2}
\end{figure}

An overview of the measurements performed on VO$_2$ (Fig. \ref{TE_VarT} and \ref{TE_VarP}) and TiO$_2$ (Fig. \ref{TE_TiO2}) under varying pressures and temperatures. The results of the TE model are shown as well as the F-N plots.

\clearpage

\section{X-ray photoemission spectroscopy}
Soft and hard X-ray photoemission spectroscopy was performed \textit{ex situ} using a dual beam PHI Quantes XPS/HAXPES scanning microprobe instrument equipped with monochromated Al (photon energy $E_{ph}^{Al} \approx$ \SI{1486.7}{\electronvolt}) and Cr anodes ($E_{ph}^{Cr} \approx$ \SI{5414.8}{\electronvolt}).
By changing both the incident photon energy as well as the photoelectron takeoff angle (between the sample surface and the analyzer entrance), we can reconstruct an approximate depth profile of the sample. 
Maximum surface sensitivity is achieved for soft X-rays and shallow takeoff angles, here using the Al anode and a takeoff angle of \SI{20}{\degree}.
Conversely, maximum bulk sensitivity is obtained for hard X-rays and normal emission, i.e., using the Cr anode and a takeoff angle of \SI{90}{\degree}.
Following the analysis of \citet{Silversmit2004}, we measure the O~1s and V~2p features in the binding energy range from 550 - \SI{505}{\electronvolt}, and reference all spectra to the O~1s peak at a binding energy of \SI{530}{\electronvolt}.
The diameter of the X-ray beam is set to \SI{100}{\um} and the source is operated in the high power setting (\SI{100}{\watt}, \SI{20}{\kilo\volt}). A low-energy electron flood gun is used for charge neutralization and low energy Ar ions are used to remove the surplus of electrons put there by the electron flood gun. A working pressure of \SI{1.3E-8}{\milli\bar} is set. The base pressure of the vacuum system is better than \SI{7E-9}{\milli\bar}.
The measurements are performed using a pass energy of \SI{140}{\electronvolt} and a step size of \SI{0.25}{\electronvolt}.

\begin{figure}[th]
	\centering
	\includegraphics[width=130mm]{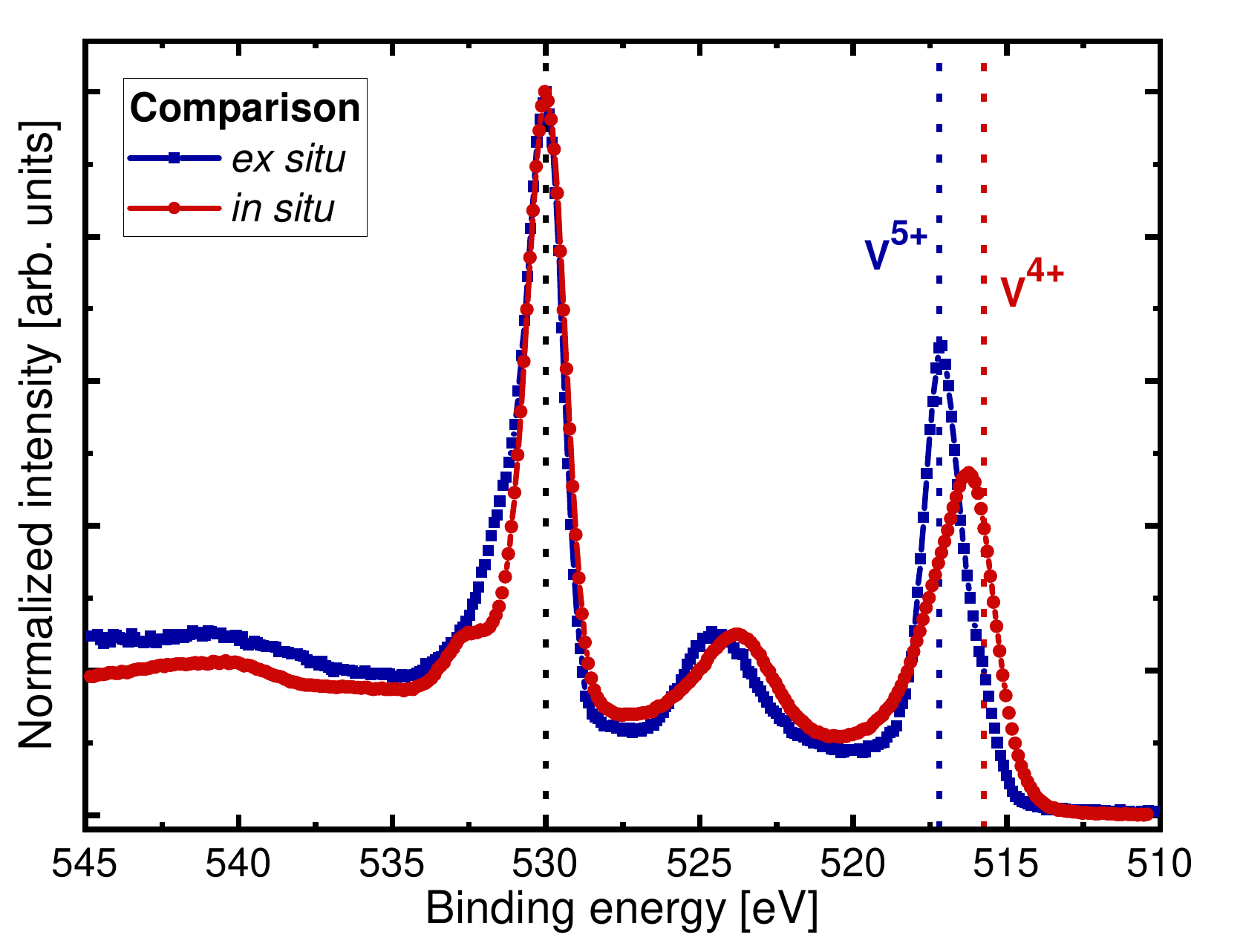}
	\caption{X-ray photoemission measurements of two \ce{VO2} thin films, one transferred \textit{in situ} to the analysis chamber, the other one \textit{ex situ}, i.e. by exposing it to the ambient air. The ratio between \ce{V^4+} and \ce{V^5+} differs, where the surface of the \textit{in situ} sample is less oxidized but is clearly not exclusively tetravalent. Both samples show a very sizable \ce{V^5+} contribution.}
	\label{Fig_XPS_insitu_exsitu}
\end{figure}

Furthermore, we observed that the \ce{V^5+} layer forms immediately and spontaneously, albeit less pronounced, even if the \ce{VO2} sample is directly transferred in ultrahigh vacuum conditions from the growth to the X-ray photoemission chamber without intermittent exposure to the ambient air.
This is displayed in Fig. \ref{Fig_XPS_insitu_exsitu}.
The \textit{in situ} XPS data has been provided by courtesy of dr. Phu Le and has been previously published \cite{Le2021}, where it was erroneously assigned as \ce{V^{4+}} exclusively.
This \textit{in situ vs. ex situ} comparative study was performed with a takeoff angle of \SI{90}{\degree} and a monochromatic Al anode source (Omicron XM 1000).

\clearpage
\section{Scanning transmission electron microscopy}
In addition to XRD and complementary to our XPS/HAXPES analysis, we also studied the \ce{VO2} thin films by scanning transmission electron microscopy (STEM) and electron energy loss spectroscopy (EELS).

\begin{figure}[tbh]
	\centering
	\includegraphics[width=100mm]{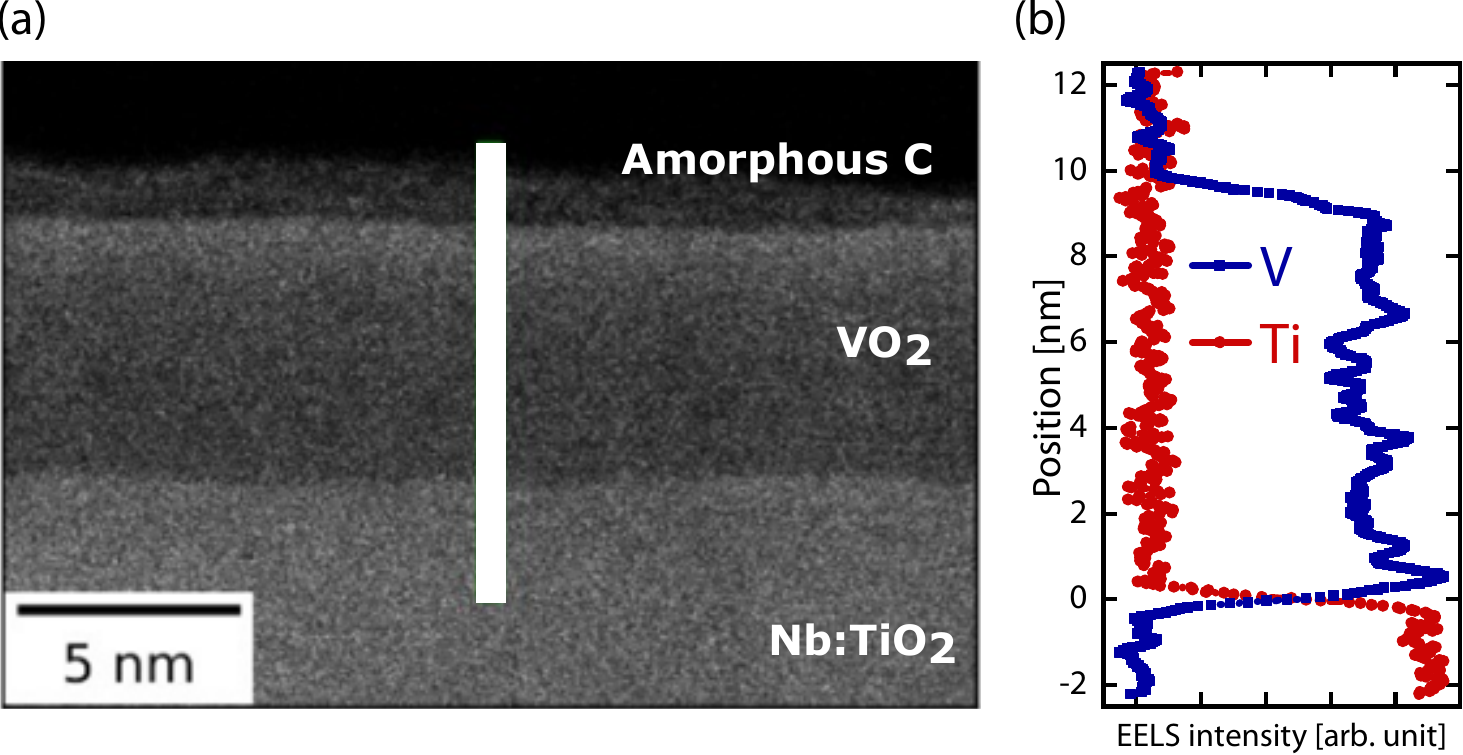}
	\caption{(a) Low magnification HAADF STEM image, (b) EELS intensity distribution for the Ti and V signals along the line indicated in (a).}
	\label{Fig_EELS}
\end{figure}

Due to the similar atomic masses of Ti and V, in high-angle annular dark-field (HAADF) imaging mode there is no clear contrast at the substrate-film interface. In Fig. 2 (d) in the main document, the atomic columns of Ti and V are virtually indiscernible, and therefore it is almost impossible to tell where the substrate ends and the film starts.
This is why we turned to monochromatic EELS to investigate the interface spectroscopically, and found indeed a sharp interface with minimal interdiffusion, see the line profile in \autoref{Fig_EELS}. To create this line profile, the intensities under the Ti and V L-edges were integrated, respectively.
The obtained EEL spectra (see Fig. 2 (c) in the main document) are in excellent agreement with our earlier work and X-ray absorption spectroscopy \cite{Le2019}.

STEM-EELS was performed at the Electron Microscopy for Materials Science (EMAT) institute at the University of Antwerp in Belgium.
Electron energy loss spectroscopy (EELS) data were acquired on a double aberration-corrected ThermoFischer Scientific Titan 80-300 electron microscope equipped with a Gatan K2 Summit camera installed at the end of a Quantum GIF. For these measurements, the microscope was operated at \SI{120}{\kV} in monochromated mode, providing an energy resolution of \SI{120}{\milli\electronvolt} with a beam current of \SI{80}{\pA}. A convergence angle of \SI{19}{\milli\radian} and a collection angle of \SI{100}{\milli\radian} were used for the acquisition. This low beam current was chosen to limit the beam-induced damage and the collection angle was maximized by the use of the EFSTEM camera lengths. An energy dispersion of \SI{0.025}{\electronvolt}~/pixel was used with an exposure time of \SI{0.2}{\second}~/pixel at \SI{100}{\percent} duty cycle of the K2 summit camera. A dose average over 32 pixels was used to further reduce beam damage.
The HAADF STEM image shown in Fig. 2 (d) in the main document was acquired at \SI{120}{\kilo\volt} with a \SI{20}{\milli\radian} convergence angle and a collection angle of \SIrange{70}{160}{\milli\radian}.

\clearpage
\section{Macroscopic transport measurements}

\begin{figure}[tbh]
	\centering
	\includegraphics[width=100mm]{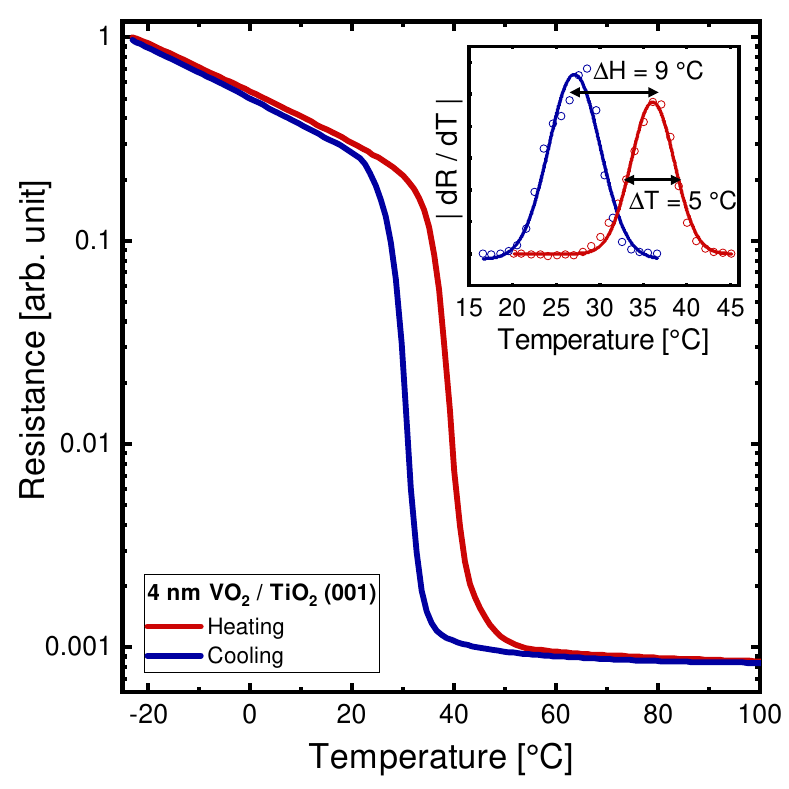}
	\caption{In-plane transport measurements as a function of temperature. Heating and cooling sweeps in red and blue, respectively. The inset shows the absolute value of the background-subtracted derivative dR/dT used to quantify the thermal hysteresis $\Delta H$ and transition sharpness $\Delta T$.}
	\label{Fig_inplane}
\end{figure}

In-plane four-point transport measurements in van der Pauw geometry reveal the well-known hysteresis curve for \ce{VO2} upon heating and cooling across the phase transition temperature. Note that for this experiment, we grew \SI{4}{\nano\meter} \ce{VO2} film on an undoped rutile \ce{TiO2} (001) substrate. Titanium-gold contacts were sputtered at the corners and connected via aluminum wire bonds to the puck of a cryogen-free Quantum Design Physical Properties Measurement System (PPMS DynaCool). In Fig. \ref{Fig_inplane}, we show the resistance as a function of temperature and observe an abrupt change of almost 3 decades across the MIT.
Apart from the amplitude of the MIT, the two other key metrics are the sharpness of the transition described by a temperature window $\Delta T$ and the width of the thermal hysteresis $\Delta H$, which can both be calculated from Gaussian fits of the derivative of the resistance ($\dv{\log R}{T}$). We find a hysteresis $\Delta H =$ \SI{9(1)}{\celsius} defined as the difference in temperatures at which the derivative curves of the heating and cooling sweeps show their respective extrema, and a transition full width half maximum of $\Delta T =$ \SI{5(1)}{\celsius} for either sweep direction.
According to Narayan and Bhosle \cite{Narayan2006} a narrow transition width is indicative of a low defect content, and the narrow hysteresis width points to a low concentration of low angle boundaries, both as expected for a coherently strained epitaxial film.

\clearpage

\bibliography{bibliography}